\documentclass[12pt,draftcls, peerreview, onecolumn]{IEEEtran/IEEEtran}
\usepackage{cite}
\usepackage{graphicx}
\usepackage{url}
\usepackage[centertags]{amsmath}
\usepackage{algorithm}
\usepackage{algorithmic}
\usepackage{amssymb}
\usepackage{multicol}
\usepackage{epsfig,color}
\newcommand{\beq}{\begin{equation}}
\newcommand{\eeq}{\end{equation}}
\newcommand{\beqn}{\begin{align}}
\newcommand{\eeqn}{\end{align}}

\begin{document}
\title{Linear Precoding of Data and Artificial Noise in Secure Massive MIMO Systems}

\author{\IEEEauthorblockN{Jun~Zhu, ~\IEEEmembership{Student Member,~IEEE}, Robert~Schober, ~\IEEEmembership{Fellow,~IEEE}, and Vijay~K.~Bhargava,~\IEEEmembership{Life Fellow,~IEEE}}\\
\IEEEauthorblockA{The University of British Columbia\\
}
\thanks{This work was presented in part at the European Wireless (EW) Conference, Barcelona, Spain, 2014, and the International Symposium on Communications, Control, and Signal Processing (ISCCSP), Athens, Greece, 2014.}}
\IEEEoverridecommandlockouts

\setcounter{page}{1}
\maketitle

\begin{abstract}
In this paper, we consider secure downlink transmission in a multi-cell massive multiple-input multiple-output (MIMO) system where the numbers of base station (BS) antennas, mobile terminals, and eavesdropper
antennas are asymptotically large. The channel state information of the eavesdropper is assumed to be unavailable at the BS and hence, linear precoding of data and artificial noise (AN) are employed for
secrecy enhancement. Four different data precoders (i.e., selfish zero-forcing (ZF)/regularized channel inversion (RCI) and collaborative ZF/RCI precoders) and three different AN precoders (i.e., random, selfish/collaborative
null-space based precoders) are investigated and the corresponding achievable ergodic secrecy rates are analyzed. Our analysis includes the effects of uplink channel estimation, pilot contamination, multi-cell interference,
and path-loss. Furthermore, to strike a balance between complexity and performance,  linear precoders that are based on matrix polynomials are proposed for both data and AN precoding. The polynomial coefficients of
the data and AN precoders are optimized respectively for minimization of the sum mean squared error of and the AN leakage to the mobile terminals in the cell of interest using tools from free probability and random matrix theory.
Our analytical and simulation results provide interesting insights for the design of secure multi-cell massive MIMO systems and reveal that the proposed polynomial data and AN precoders closely approach the performance of selfish RCI
data and null-space based AN precoders, respectively.
\end{abstract}


\section{Introduction}
Massive multiple-input multiple-output (MIMO) systems employing simple linear precoding and combining schemes offer significant performance gains in terms of bandwidth, power,
and energy efficiency compared to conventional multiuser MIMO systems as impairments such as fading, noise, and interference are averaged out for very large numbers of base
station (BS) antennas \cite{survey,noncooperative,energyspectraleff}. Furthermore, in time-division duplex (TDD) systems, channel reciprocity can be exploited to estimate the
downlink channels via uplink training so that the training overhead scales only linearly with the number of users and is independent of the number of BS antennas \cite{noncooperative}.
However, if the pilot sequences employed in different cells are not orthogonal, so-called pilot contamination occurs and impairs the channel estimates, which ultimately limits the achievable
performance of massive MIMO systems \cite{noncooperative,pilotcontam}.

Since secrecy and privacy are critical concerns for the design of future communication systems \cite{physurvey}, it is of interest to
investigate how the large number of spatial degrees of freedom in massive MIMO systems can be exploited for secrecy enhancement \cite{khisti,hassibi2}. If the eavesdropper (Eve)
remains passive to hide its existence, neither the transmitter (Alice) nor the legitimate receiver (Bob) will be able to learn Eve's channel state information (CSI). In this situation, it is advantageous
to inject artificial noise (AN) at the transmitter to degrade Eve's channel and to use linear precoding to avoid impairment to Bob's channel as was shown in \cite{negi}-\cite{robsec} and \cite{qos}, \cite{masksec} for single user
and single-cell multiuser systems, respectively. However, in multi-cell massive MIMO systems, multi-cell interference and pilot contamination will hamper Alice's ability to degrade Eve's channel and to protect Bob's
channel. This problem was studied first in \cite{zhu} for simple matched-filter (MF) data precoding and null-space (NS) and random AN precoding. However, it is well known that MF data precoding
suffers from a large loss in the achievable information rate compared to other linear data precoders such as zero-forcing (ZF) and regularized channel inversion (RCI) precoders as the number of mobile
terminals (MTs) increases \cite{uldl}. Since it is expected that this loss in information rate also translates into a loss in secrecy rate, studying the secrecy performance of ZF and RCI data precoders in massive
MIMO systems is of interest. Furthermore, while NS AN precoding was shown to achieve a better performance compared to random AN precoding \cite{zhu}, it also entails a much higher complexity. Similarly,
the improved performance of ZF and RCI data precoding compared to MF data precoding comes at the expense of a higher complexity. Hence, the design of novel data and AN precoders which allow a flexible
tradeoff between complexity and secrecy performance is  desirable.

Related work on physical layer security in massive MIMO systems includes \cite{goldsmith} where the authors use the channel between Alice and Bob as secrete key and show that the complexity required by Eve
to decode Alice's message is at least of the same order as a worst-case lattice problem. Physical layer security in a downlink multi-cell MIMO system was considered in \cite{multisec}-\cite{largerci}. However, unlike
our work, perfect knowledge of Eve's channel was assumed, AN injection was not considered, and pilot contamination was not taken into account.  Furthermore, ZF and RCI data precoding were analyzed in the
large system limit in \cite{zf,mmse}. However, neither pilot contamination nor AN were taken into account and the secrecy rate was not analyzed. {Using a concept that was originally conceived for code
division multiple access (CDMA) uplink systems in \cite{polypio} and later extended to MIMO systems in \cite{polymmimo}, reduced complexity linear data precoders that are based on matrix polynomials were
investigated for use in massive MIMO systems in \cite{poly}-\cite{poly2}. However, \cite{poly}-\cite{poly2} did not take into account the effect of AN leakage for precoder design and did not study the secrecy performance.}
Hence, the results presented in \cite{goldsmith}-\cite{poly2} are not directly applicable to the system studied in this paper.

In this paper, we consider secure downlink transmission in a multi-cell massive MIMO system employing linear data and AN precoding in the presence of a passive multi-antenna eavesdropper. We study the achievable
ergodic secrecy rate of such systems for different linear precoding schemes taking into account the effects of uplink channel estimation, pilot contamination, multi-cell interference, and path-loss.
{The main contributions of this paper are summarized as follows:
\begin{itemize}
  \item We study the performance-complexity tradeoff of selfish and collaborative data and AN precoders. Selfish precoders require only the CSI of the MTs in the local cell but cause inter-cell interference and inter-cell AN leakage.
  In contrast, collaborative precoders require the CSI between the local BS and the MTs in all cells, but reduce inter-cell interference and inter-cell AN leakage. However, since the additional CSI required for the collaborative precoders
  can be estimated directly by the local BS, the additional overhead and complexity incurred compared to selfish precoders is limited.
  \item We derive novel closed-form expressions for the asymptotic ergodic secrecy rate which facilitate the performance comparison of different combinations of linear data precoders (i.e., MF, selfish and collaborative ZF/RCI)
  and AN precoders (i.e., random, selfish and collaborative NS), and provide significant insight for system design and optimization.
  \item In order to avoid the computational complexity and potential stability issues in fixed point implementations entailed by the large-scale matrix inversions required for ZF and RCI data precoding and NS AN precoding,
  we propose polynomial (POLY) data and AN precoders and optimize their coefficients. Unlike \cite{muller} and \cite{poly2}, which considered polynomial data precoders for massive MIMO systems without AN generation, we use
 free probability theory \cite{poly,free} to obtain the POLY coefficients. This allows us to express the POLY coefficients as simple functions of the channel and system parameters. Simulation results reveal
 that these precoders are able to closely approach the performance of selfish RCI data and NS AN precoders, respectively.
\end{itemize}
}
The remainder of this paper is organized as follows. In Section \ref{s2}, we outline the considered system model and review some basic results from \cite{zhu}. In Sections \ref{s3} and \ref{s4}, the considered linear data and AN precoders
are investigated, respectively. In Section \ref{s5}, the ergodic secrecy rates of different linear precoders are compared analytically for a simple path-loss model. Simulation and numerical results are presented in Section \ref{s6}, and some
conclusions are drawn in Section \ref{s7}.

\textit{Notation:} Superscripts $T$ and $H$ stand for the transpose and conjugate transpose, respectively. ${\bf I}_N$ is the $N$-dimensional identity matrix. The expectation operation and the variance of a random variable are denoted
by $\mathbb{E}[\cdot]$ and ${\rm var}[\cdot]$, respectively. ${\rm diag}\{{\bf x}\}$ denotes a diagonal matrix with the elements of vector ${\bf x}$ on the main diagonal. ${\rm tr}\{\cdot\}$ and ${\rm rank}\{\cdot\}$ denote trace and rank of a
matrix, respectively. $\mathbb{C}^{m\times n}$ represents the space of all $m \times n$ matrices with complex-valued elements. ${\bf x} \sim \mathbb{CN}({\bf 0}_N, \boldsymbol{\Sigma})$ denotes a circularly symmetric complex Gaussian
vector ${\bf x}\in \mathbb{C}^{N \times 1}$ with zero mean and covariance matrix $\boldsymbol{\Sigma}$. $[{\bf A}]_{kl}$ denotes the element in the $k^{\rm th}$ row and $l^{\rm th}$ column of matrix ${\bf A}$, and $[x]^+=\max\{x,0\}$.
\section{System Model and Preliminaries} \label{s2}
In this section, we introduce the considered system model as well as the adopted channel estimation scheme, and review some ergodic secrecy rate results.
\subsection{System Model}\label{s2a}
We consider the downlink of a multi-cell massive MIMO system with $M$ cells and a frequency reuse factor of one, i.e., all BSs use the same spectrum. Each cell includes one $N_T$-antenna BS, $K \le N_T$ single-antenna MTs, and
potentially an $N_E$-antenna eavesdropper. The eavesdroppers try to hide their existence and hence remain passive. As a result, the BSs cannot estimate the eavesdroppers' CSI. To overcome this limitation, each BS generates AN
to mask its information-carrying signal and to prevent eavesdropping \cite{negi}. In the following, the $k^{\rm th}$ MT, $k=1,\ldots, K$, in the $n^{\rm th}$ cell, $n=1,\ldots, M$, is the MT of interest and we assume that an eavesdropper tries to
decode the signal intended for this MT. {We note that neither the BSs nor the MTs are assumed to know which MT is targeted by the eavesdropper.} The signal vector, ${\bf x}_n \in \mathbb{C}^{N_T \times 1}$, transmitted by
the BS in the $n^{\rm th}$ cell (also referred to as the $n^{\rm th}$ BS in the following) is given by
\begin{equation}
{\bf x}_n = \sqrt{p} {\bf F}_n {\bf s}_n + \sqrt{q} {\bf A}_n {\bf z}_n,
\label{eq1}
\end{equation}
where ${\bf s}_n \sim \mathbb{CN}({\bf 0}_{K},  {\bf I}_{K})$ and ${\bf z}_n  \sim \mathbb{CN}({\bf 0}_{N_T},  {\bf I}_{N_T})$ denote the data and AN vectors for the $K$ MTs in the $n^{\rm th}$ cell, respectively.
${\bf F}_n =[{\bf f}_{n1},\cdots,{\bf f}_{nK}] \in \mathbb{C}^{N_T \times K}$ and ${\bf A}_n =[{\bf a}_{n1},\cdots,{\bf a}_{nN_T}]\in \mathbb{C}^{N_T \times N_T}$ are the data and AN precoding matrices, respectively,
and the efficient design of these matrices is the main scope of this paper.  {Thereby, the structure of both types of precoding matrices does not depend on which MT is targeted by the eavesdropper.}
The AN precoding matrix ${\bf A}_n$ has rank $L={\rm rank}\{{\bf A}_n\}\le N_T$, i.e., $L$ dimensions of the $N_T$-dimensional signal space
spanned by the $N_T$ BS antennas are exploited for jamming of the eavesdropper. The data and AN precoding matrices are normalized as ${\rm tr}\{{\bf F}^H_n{\bf F}_n\}=K$ and ${\rm tr}\{{\bf A}^H_n{\bf A}_n\}=L$, i.e.,
their average power per dimension is one. The average powers $p$ and $q$ allocated to the information-carrying signal for each MT and each AN signal, respectively, can be written as $p=\frac{\phi P_T}{K}$ and $q=\frac{(1-\phi)P_T}{L}$,
where $P_T$ is the total transmit power and $\phi \in (0,1]$ is a power allocation factor which can be optimized. For the sake of clarity, in this paper, we assume that all cells utilize the same value of $\phi$.

The vectors collecting the received signals at the $K$ MTs and the $N_E$ antennas of the eavesdropper in the $n^{\rm th}$ cell are given by
\begin{equation}
{\bf y}_n=\sum_{m=1}^M {\bf G}_{mn} {\bf x}_m+{\bf n}_n \qquad {\rm and} \qquad  {\bf y}_{E}=\sum_{m=1}^M {\bf G}_{mE} {\bf x}_m+{\bf n}_E,
\label{eq2}
\end{equation}
respectively, with Gaussian noise vectors ${\bf n}_n \in \mathbb{CN}({\bf 0}_{K},\sigma^2_n {\bf I}_K)$ and ${\bf n}_E \in \mathbb{CN}({\bf 0}_{N_E},\sigma^2_E {\bf I}_{N_E})$, where $\sigma_n^2$ and $\sigma_E^2$ denote the noise variances
at one MT and one eavesdropper receive antenna, respectively. Furthermore, ${\bf G}_{mn}={\bf D}^{1/2}_{mn} {\bf H}_{mn} \in \mathbb{C}^{K \times N_T}$ and ${\bf G}_{mE} =\sqrt{\beta_{mE}}{\bf H}_{mE}\in \mathbb{C}^{N_E \times N_T}$
are the matrices modeling the channels from the $m^{\rm th}$ BS to the $K$ MTs and the eavesdropper in the $n^{\rm th}$ cell, respectively. Thereby, ${\bf D}_{mn}={\rm diag}\{\beta^1_{mn},\ldots,\beta^K_{mn}\}$
and $\beta_{mE}$ represent the path-losses from the $m^{\rm th}$ BS to the $K$ MTs and the eavesdropper in the $n^{\rm th}$ cell, respectively. Matrix ${\bf H}_{mn} \in \mathbb{C}^{K \times N_T}$, with row vector
${\bf h}^k_{mn} \in \mathbb{C}^{1 \times N_T}$ in the $k^{\rm th}$ row, and matrix ${\bf H}_{mE}\in \mathbb{C}^{N_E \times N_T}$ represent the corresponding small-scale fading components. Their elements are modeled as mutually independent and
identically distributed (i.i.d.) complex Gaussian random variables (RVs) with zero mean and unit variance.

For the design of the data and noise precoders, we consider two different approaches:  \textit{Selfish} designs and  \textit{collaborative} designs.
For the selfish designs, each BS designs its precoders only based on the estimate of the CSI in its own cell, ${\bf G}_{nn}$, and without regard for the interference and the AN
it causes to other cells. In contrast, for the collaborative designs, each BS designs its precoders based on the estimates of the CSI to the MTs in all cells, ${\bf G}_{mn}$, $m=1,
\ldots, M$, in an effort to avoid excessive interference and AN to other cells. Although collaborative designs introduce more channel estimation overhead {at the BS},
they may not always outperform selfish designs because of the imperfection of the CSI and the limited number of spatial degrees of freedom available for precoder design.
\subsection{Channel Estimation and Pilot Contamination}\label{s2b}
As is customary for massive MIMO systems, we assume that the downlink and uplink channels are reciprocal and the CSI is estimated in an uplink training phase \cite{survey}-\cite{pilotcontam}. To this end, all MTs emit pilot sequences of length
$\tau\ge K$ and with pilot symbol power $p_\tau$. We assume that the pilot sequences of the $K$ MTs in a given cell are mutually orthogonal but the same pilot sequences are used in all cells. This gives rise to so-called pilot contamination
\cite{survey}-\cite{pilotcontam}. Furthermore, we assume that the path-loss information changes on a much slower time scale than the small-scale fading. Hence, the path-loss matrices ${\bf D}_{nm}$, $m=1,\ldots,M$,
can be estimated perfectly and are assumed to be known at the BS for minimum mean-square error (MMSE) estimation of the small-scale fading gains \cite{pilotcontam}. At the $n^{\rm th}$ BS, the small-scale fading vector to the
$k^{\rm th}$ MT in the $m^{\rm th}$ cell, ${\bf h}^k_{nm}$, can be expressed as
\begin{equation}
{\bf h}^k_{nm}=\hat{\bf h}^k_{nm}+\tilde{\bf h}^k_{nm},
\label{eq3}
\end{equation}
where the estimate $\hat{\bf h}^k_{nm}$ and the estimation error $\tilde{\bf h}^k_{nm}$ are mutually independent and can be statistically characterized as $\hat{\bf h}^k_{nm} \sim \mathbb{CN}({\bf 0}_{N_T},\frac{p_\tau \tau \beta^k_{nm}}{1+p_\tau
\tau \sum_{l=1}^M \beta^k_{nl}}{\bf I}_{N_T})$ and $\tilde{\bf h}^k_{nm}\sim \mathbb{CN}({\bf 0}_{N_T},\frac{1+p_\tau \tau \sum_{l \neq m}^M \beta^k_{nl}}{1+p_\tau \tau \sum_{l=1}^M \beta^k_{nl}}{\bf I}_{N_T})$, respectively, cf.~\cite{zhu}.
For future reference, we collect the estimates and the estimation errors at the $n^{\rm th}$ BS corresponding to all $K$ MTs in the $m^{\rm th}$ cell in matrices $\hat{\bf H}_{nm} =[ (\hat{{\bf h}}^{1}_{nm})^T, \ldots, (\hat{\bf h}^K_{nm} )^T]^T
\in \mathbb{C}^{K \times N_T}$ and $\tilde{\bf H}_{nm}=[ (\tilde{{\bf h}}^{1}_{nm})^T, \ldots, (\tilde{{\bf h}}^{K}_{nm} )^T]^T \in \mathbb{C}^{K \times N_T}$, respectively.
\subsection{Ergodic Secrecy Rate}\label{s2c}
The performance metric adopted in this paper is the ergodic secrecy rate \cite{hassibi2}. In this section, we review some results for the ergodic secrecy rate in multi-cell massive MIMO systems employing linear data and AN precoding
from \cite{zhu}, as these results will be needed throughout this paper. Combining (\ref{eq1}) and (\ref{eq2}) we observe that the downlink channel comprising the BS, the $k^{\rm th}$ MT, and the
eavesdropper in the $n^{\rm th}$ cell is an instance of a multiple-input, single-output, multi-eavesdropper (MISOME) wiretap channel \cite{khisti}. Hence, the achievable secrecy rate of the $k^{\rm th}$ MT in the $n^{\rm th}$
cell is bounded by the difference of the capacities of the channel between the BS and the MT and the channel between the BS and the eavesdropper, see \cite[Lemma 1]{zhu}, \cite[Lemma 2]{massivesec}.
Thus, a lower bound on the ergodic secrecy rate of the $k^{\rm th}$ MT in the $n^{\rm th}$ cell is given by \cite{zhu}
\begin{equation}
\label{secnk}
R_{nk}^{\rm sec}=[R_{nk}-C^{\rm eve}_{nk}]^+, {\color{blue} k=1,\ldots,K,}
\end{equation}
where $R_{nk}$ denotes an achievable rate of the $k^{\rm th}$ MT in the $n^{\rm th}$ cell and $C^{\rm eve}_{nk}$ denotes the ergodic capacity of the channel between the BS and the eavesdropper. In order to obtain a tractable lower bound on the
ergodic secrecy rate, we lower bound the achievable rate of the MT as $R_{nk}=\log_2(1+\gamma_{nk})$ with signal-to-interference-and-noise ratio (SINR) \cite[Eq.~(10)]{zhu}
\begin{equation}
\label{Rnk}
\gamma_{nk}=\frac{|\mathbb{E}[\sqrt{\beta^k_{nn} p} {\bf h}^k_{nn} {\bf f}_{nk}]|^2}{{{\rm var}[\sqrt{\beta^k_{nn} p} {\bf h}^k_{nn} {\bf f}_{nk}]}+\sum\limits^M_{m=1} \sum\limits_{i=1}^{N_t} \mathbb{E}[|\sqrt{\beta^k_{mn} q}{\bf h}^k_{mn} {\bf a}_{mi}|^2]+
\sum\limits_{\{m,l\} \neq \{n,k\}} \mathbb{E}[|\sqrt{ \beta^k_{mn} p} {\bf h}^k_{mn} {\bf f}_{ml}|^2]+1}.
\end{equation}
Furthermore, we make the pessimistic assumption that the eavesdropper is able to cancel the received signals of all in-cell and out-of-cell MTs except the signal intended for the MT of interest.
{This leads to an upper bound for the eavesdropper's capacity, and consequently, to a lower bound for the ergodic secrecy rate.}\footnote{{This lower bound is achievable if the eavesdropper has access to the
data of all interfering in-cell and out-of-cell MTs, which might be the case e.g.~if the interfering MTs cooperate with the eavesdropper.}} Hence, the ergodic capacity of the eavesdropper is given by \cite[Eq.~(7)]{zhu}
\begin{equation}
\label{Ceve}
C^{\rm eve}_{nk} = \mathbb{E}\bigg[\log_2 \left(1+p {\bf f}^H_{nk} {\bf G}^{H}_{nE} {\bf X}^{-1}{\bf G}_{nE} {\bf f}_{nk}\right)\bigg],
\end{equation}
where ${\bf X}=q \sum_{m=1}^M {\bf G}_{mE} {\bf A}_{m} {\bf A}^H_{m} {\bf G}^H_{mE} \in \mathbb{C}^{N_T\times N_T}$ denotes the noise correlation matrix at the eavesdropper under the worst-case assumption that the
receiver noise at the eavesdropper is negligible, i.e., $\sigma_E^2\to 0$. Denoting the normalized number of eavesdropper antennas  by $\alpha=N_E/N_T$, a necessary condition
for the invertibility of matrix ${\bf X}$ is $\alpha\le ML/N_T$. Hence, a non-zero secrecy rate can only be achieved if this condition is met. Consequently, a larger $L$ implies that the BS is able to tolerate more
eavesdropper antennas.

If ${\bf H}_{nE}{\bf f}_{nk}$ and matrix ${\bf X}$ are statistically independent, which in turn means for the data and AN precoders that vector ${\bf f}_{nk}$ and the subspace spanned by the columns of ${\bf A}_n$ are mutually orthogonal,
a simple and tight upper bound on (\ref{Ceve}) can be obtained. Since any efficient data/AN precoder pair has to keep the AN self-interference at the desired MT small, this orthogonality condition holds at least approximately in practice.
In this case, for $\alpha<a^2L/(c N_T)$ { and $N_T\to \infty$}, where $a=1+\sum_{m \neq n}^M \beta_{mE}/\beta_{nE}$ and $c=1+\sum_{m \neq n}^M (\beta_{mE}/\beta_{nE})^2$, a simple and tight upper bound for $C^{\rm eve}_{nk}$ is given by
\cite[Theorem 1]{zhu}
\begin{equation}
\label{Cup}
C^{\rm eve}_{nk} \leq \log_2 \left(1+\frac{\alpha p}{ aq L/N_T -c \alpha q/a}\right)= \log_2 \left(1+\frac{\alpha \phi}{\beta(1-\phi)( a-c \alpha N_T/(L a))}\right).
\end{equation}

For $M=1$, we have $a^2/c=M=1$, i.e., the bound in (\ref{Cup}) is applicable in the entire range of $\alpha$ where $C^{\rm eve}_{nk}$ in (\ref{Ceve}) is finite. For $M>1$, we have $a^2/c\le M$, i.e.,
the bound is not applicable for $La^2/(c N_T)\le \alpha \le ML/N_T$. However, for strong inter-cell interference, we have $\beta_{mE}\approx \beta_{nE}$ and $a^2/c\approx M$, i.e., the
bound is applicable for all $\alpha$ for which $C^{\rm eve}_{nk}$ in (\ref{Ceve}) is finite. On the other hand, for weak inter-cell interference, we have $\beta_{mE} \ll \beta_{nE}$, and matrix ${\bf X}$
will be ill-conditioned for $L/N_T\le \alpha\le ML/N_T$ and $C^{\rm eve}_{nk}$ will become very large. Hence, the bound is again applicable for the values of $\alpha$ (i.e., $0\le \alpha \le L/N_T$), for which
$C^{\rm eve}_{nk}$ in (\ref{Ceve}) assumes practically relevant values. More generally, \cite[Figs.~2-4]{zhu} and Section \ref{s6} suggest that{, for $N_T\to\infty$,} (\ref{Cup}) is applicable and tight for all values of $\alpha$
which permit a non-vanishing secrecy rate.

Combining (\ref{secnk}), (\ref{Rnk}), and (\ref{Cup}), we obtain a tight and tractable lower bound on the secrecy rate \cite{zhu}. It is noteworthy that the upper bound on the capacity of the eavesdropper
in (\ref{Cup}) is only affected by the dimensionality of the AN precoder, $L$, but not by the exact structures of ${\bf A}_n$ and ${\bf F}_n$, as long as ${\bf f}_{nk}$ and the subspace spanned by the columns
of ${\bf A}_n$ are orthogonal. On the other hand, the achievable rate of the MT in (\ref{Rnk}) is affected by both the data and the AN precoders. In the following two sections, we analyze the impact of the
most important existing data and AN precoder designs on the achievable rate $R_{nk}$ as $N_T\to\infty$, respectively, and propose novel low-complexity data and AN precoders that are based on a
polynomial matrix expansion.
\section{Linear Data Precoders for Secure Massive MIMO}\label{s3}
In this section, we analyze the achievable rate of selfish and collaborative ZF/RCI data precoding, respectively, and develop a novel POLY data precoder. {In contrast to existing analyses and designs of data precoders
for massive MIMO, e.g.~\cite{zf,mmse}, \cite{poly}-\cite{poly2}, the results presented in this  section account for the effect of AN leakage, which is only present if AN is injected at the BS for secrecy enhancement.}
We are interested in the asymptotic regime where $K, N_T\to\infty$ but $\beta =K/N_T$ and $\alpha=N_E/N_T$ are finite.
\subsection{Analysis of Existing Data Precoders}\label{s3a}
For $N_T\to \infty$, analyzing the achievable rate is equivalent to analyzing the SINR in (\ref{Rnk}). Thereby, the effect of the AN precoder can be captured by the term
\begin{equation}
Q=\sum^M_{m=1} \sum_{i=1}^{N_t} \mathbb{E}[|\sqrt{\beta^k_{mn} }{\bf h}^k_{mn} {\bf a}_{mi}|^2]= \sum^M_{m=1} \beta^k_{mn} \mathbb{E}[{\bf h}^k_{mn} {\bf A}_{m} {\bf A}^H_{m} ({\bf h}^k_{mn})^H]
\label{eq8}
\end{equation}
in the denominator of (\ref{Rnk}), which represents the inter-cell and intra-cell AN leakage. This term is assumed to be given in this section and will be analyzed in detail for different AN precoders in Section \ref{s4}.
\subsubsection{Selfish ZF/RCI Data Precoding}
The selfish RCI (SRCI) data precoder for the $n^{\rm th}$ cell is given by
\begin{equation}
\label{srci}
{\bf F}_{n}=\gamma_1 {\bf L}_{nn} \hat{\bf H}^H_{nn},
\end{equation}
where ${\bf L}_{nn}=(\hat{\bf H}^H_{nn} \hat{\bf H}_{nn}+\kappa_1 {\bf I}_{N_T})^{-1}$, $\gamma_1$ is a scalar normalization constant, and $\kappa_1$ is a regularization constant.
In the following proposition, we provide the resulting SINR of the $k^{\rm th}$ MT in the $n^{\rm th}$ cell.

\textit{Proposition 1}: For SRCI data precoding, the received SINR at the $k^{\rm th}$ MT in the $n^{\rm th}$ cell is given by
\begin{equation}
\label{gammalmmseNpc}
\gamma^{{\rm SRCI}}_{nk}=\frac{1}{\frac{\hat{\Gamma}_{\rm SRCI}+(1+\mathcal{G}(\beta,\kappa_1))^2}{\mathcal{G}(\beta,\kappa_1)\left(\hat{\Gamma}_{\rm SRCI} +\frac{\hat{\Gamma}_{\rm SRCI} \kappa_1}{\beta}(1+\mathcal{G}(\beta,\kappa_1))^2\right)}+\sum_{m \neq n} \beta^k_{mn}/\beta^k_{nn}},
\end{equation}
where
\begin{equation}
\label{g}
\mathcal{G}(\beta,\kappa_1)=\frac{1}{2} \bigg[\sqrt{\frac{(1-\beta)^2}{\kappa_1^2}+\frac{2(1+\beta)}{\kappa_1}+1}+\frac{1-\beta}{\kappa_1}-1\bigg],
\end{equation}
and $\hat{\Gamma}_{\rm SRCI}=\frac{\Gamma_{\rm SRCI} \theta_{nk}}{\Gamma_{\rm SRCI} \vartheta_{nk}+1}$ with ${\Gamma}_{\rm SRCI}=\frac{\beta^k_{nn} K}{\sum_{m \neq n}^M \sum_{l \neq k} \beta^k_{mn} +
\eta Q+\frac{K}{\phi P_T}}$, $\theta_{mk}=\frac{p_\tau \tau (\beta^k_{mn})^2}{1+p_\tau \tau \sum_{l=1}^M \beta^k_{ml}}$, $\vartheta_{mk}=\beta^k_{mn}$ $\times \frac{1+p_\tau \tau \sum_{l \neq m}^M
\beta^k_{ml}}{1+p_\tau \tau \sum_{l=1}^M \beta^k_{ml}}$, and $\eta=q/p$.
\begin{IEEEproof}
Please refer to Appendix A.
\end{IEEEproof}
Regularization constant $\kappa_1$ can be optimized for maximization of the lower bound on the secrecy rate in (\ref{secnk}), which is equivalent to maximizing the SINR in (\ref{gammalmmseNpc}). Setting the derivative
of $\gamma^{{\rm SRCI}}_{nk}$ with respect to $\kappa_1$ to zero, the optimal regularization parameter is found as $\kappa_{1,{\rm opt}}=\beta/\hat{\Gamma}_{\rm SRCI}$, and the corresponding maximum SINR is given
by
\begin{equation}
\label{srci-opt}
\gamma^{{\rm SRCI}}_{nk}=\frac{1}{1/\mathcal{G}(\beta,\kappa_{1,{\rm opt}})+\sum_{m \neq n} \beta^k_{mn}/\beta^k_{nn}}.
\end{equation}
On the other hand, for $\kappa_1 \to 0$, the SRCI data precoder in (\ref{srci}) reduces to the selfish ZF (SZF) data precoder. The corresponding received SINR is provided in the following corollary.

\textit{Corollary 1}: Assuming $\beta\le 1$, for SZF data precoding, the received SINR at the $k^{\rm th}$ MT in the $n^{\rm th}$ cell is given by
\begin{equation}
\label{szf}
\gamma^{{\rm SZF}}_{nk}=\frac{1}{\frac{\beta}{(1-\beta)\hat{\Gamma}_{\rm SRCI}}+\sum_{m \neq n}\beta^k_{mn}/\beta^k_{nn}}.
\end{equation}
\begin{IEEEproof}
$\gamma^{{\rm SZF}}_{nk}$ in (\ref{szf}) can be obtained from (\ref{gammalmmseNpc}) as $\gamma^{{\rm SZF}}_{nk}=\lim_{\kappa_1\to 0} \gamma^{{\rm SRCI}}_{nk}$.
\end{IEEEproof}
\subsubsection{Collaborative ZF/RCI Precoding}
The collaborative RCI (CRCI) precoder for the $n^{\rm th}$ cell is given by
\begin{equation}
\label{crci}
{\bf F}_{n}=\gamma_2 {\bf L}_n \hat{\bf H}^H_{nn},
\end{equation}
where ${\bf L}_n=(\hat{\bf H}^H_n \hat{\bf H}_n+\kappa_2 {\bf I}_{N_T})^{-1}$ with $\hat{\bf H}_n=[ \hat{\bf H}_{n1}^T\ldots \hat{\bf H}_{nM}^T ]^T \in \mathbb{C}^{MK \times N_T}$,  $\gamma_2$ is a normalization
constant, and $\kappa_2$ is a regularization constant. The corresponding SINR of the $k^{\rm th}$ MT in the $n^{\rm th}$ cell is provided in the following proposition.

\textit{Proposition 2}: For CRCI data precoding, the received SINR at the $k^{\rm th}$ MT in the $n^{\rm th}$ cell is given by
\begin{equation}
\label{gammaCRCINpc}
\gamma^{{\rm CRCI}}_{nk} = \frac{1}{\frac{\hat{\Gamma}_{\rm CRCI}+(1+\mathcal{G}(M \beta,\kappa_2))^2}{\mathcal{G}(M \beta,\kappa_2)\left(\hat{\Gamma}_{\rm CRCI} +\frac{\hat{\Gamma}_{\rm CRCI} \kappa_2}{\beta}(1+\mathcal{G}(M \beta,\kappa_2))^2\right)}+\sum_{m \neq n} \beta^k_{mn}/\beta^k_{nn}},
\end{equation}
where $\hat{\Gamma}_{\rm CRCI}=\frac{\Gamma_{\rm CRCI} \theta_{nk}}{\Gamma_{\rm CRCI} \vartheta_{nk}+1}$ with $\Gamma_{\rm CRCI}=\frac{\beta^k_{nn} K}{\eta Q+\frac{K}{\phi P_T}}$.
\begin{IEEEproof}
The proof is similar to that for the SINR for the SRCI data precoder given in Appendix A and omitted here for brevity.
\end{IEEEproof}

Furthermore, the optimal regularization constant maximizing the SINR (and thus the secrecy rate) in (\ref{gammaCRCINpc}) is obtained as $\kappa_{2,{\rm opt}}=M \beta/\hat{\Gamma}_{\rm CRCI}$, and the
corresponding maximum SINR is given by
\begin{equation}
\label{crci-opt}
\gamma^{{\rm CRCI}}_{nk}=\frac{1}{1/\mathcal{G}(M \beta,\kappa_{2,{\rm opt}})+\sum_{m \neq n} \beta^k_{mn}/\beta^k_{nn}}.
\end{equation}

On the other hand, for $\kappa_2 \to 0$, the CRCI precoder in (\ref{crci}) reduces to the collaborative ZF (CZF) precoder. The corresponding received SINR is provided in the following corollary.

\textit{Corollary 2}: Assuming $\beta\le 1/M$, for CZF data precoding, the received SINR at the $k^{\rm th}$ MT in the $n^{\rm th}$ cell is given by

\begin{equation}
\label{czf}
\gamma^{{\rm CZF}}_{nk}=\frac{1}{\frac{M\beta}{(1-M\beta)\hat{\Gamma}_{\rm CRCI}}+\sum_{m \neq n}\beta^k_{mn}/\beta^k_{nn}}.
\end{equation}

\begin{IEEEproof}
$\gamma^{{\rm CZF}}_{nk}$ in (\ref{czf}) is obtained by letting $\kappa_2\to 0$ in (\ref{gammaCRCINpc}).
\end{IEEEproof}

\textit{Remark 1:} Selfish data precoders require estimation of in-cell CSI, i.e., $\hat{\bf H}_{nn}$, only. In contrast, collaborative data precoders require estimation of both in-cell and inter-cell CSI at the BS, i.e., $\hat{\bf H}_n$. Furthermore,
since collaborative data precoders attempt to avoid interference not only to in-cell users but also to out-of-cell users, more BS antennas are needed to achieve high performance. This is evident from Corollaries 1 and 2, which
reveal that $N_T>K$ and $N_T>MK$ are necessary for SZF and CZF data precoding, respectively. On the other hand, if successful, trying to avoid out-of-cell interference is beneficial for the overall
performance. Hence,  whether selfish or collaborative precoders are preferable depends on the parameters of the considered system, cf.~Sections \ref{s5} and \ref{s6}.
\subsection{Polynomial Data Precoder}\label{s3b}
The RCI and ZF data precoders introduced in the previous section achieve a higher performance than simple MF data precoding \cite{zhu}. However, they require a matrix inversion which entails a high computational complexity
for the large values of $K$ and $N_T$ desired in massive MIMO. Hence, in this section, we propose a low-complexity POLY data precoder which avoids the matrix inversion. As the goal is a low-complexity design, we
focus on selfish POLY precoders, although the extension to collaborative designs is possible.

The proposed POLY precoder, ${\bf F}_n$, for the $n^{\rm th}$ BS can be expressed as
\begin{equation}
\label{W}
{\bf F}_n=\frac{1}{\sqrt{N_T}}\hat{\bf \overline{H}}^H_{nn} \sum_{i=0}^{\cal I} \mu_i \left(\hat{\bf \overline{H}}_{nn} \hat{\bf \overline{H}}^H_{nn}\right)^i,
\end{equation}
where $\hat{\bf \overline{H}}_{nn}=\frac{1}{\sqrt{N_T}} \hat{\bf H}_{nn}$, and $\boldsymbol{\mu}=[\mu_0,\ldots,\mu_{\cal I}]^T$ are the real-valued coefficients of the precoder matrix polynomial, which have to be optimized.
In the following, we show that, for $K,N_T\to\infty$,  the optimum coefficients $\boldsymbol{\mu}$ do not depend on the instantaneous channel estimates but are constant and can be determined by exploiting results from
{free probability \cite{free}} and random matrix theory \cite{rmt}. To this end, we define the asymptotic average mean-square error (MSE) of the users in the $n^{\rm th}$ cell  as ${\rm mse}_n = \lim_{K \to
\infty }\frac{1}{K}\mathbb{E}\left[\|{\bf e}_n\|^2\right]$ with error vector
\begin{equation}
\label{e}
{\bf e}_n = \varsigma {\bf y}_n-{\bf s}_n= \varsigma({\bf G}_{nn} (\sqrt{p}{\bf F}_n {\bf s}_n+\sqrt{q}{\bf A}_n {\bf z}_n) +\tilde{\bf n}_n)-{\bf s}_n,
\end{equation}
where $\tilde{\bf n}_n= \sum_{m\ne n} {\bf G}_{mn}{\bf x}_m+{\bf n}_n$ includes Gaussian noise, inter-cell interference, and inter-cell AN leakage. Furthermore, $\varsigma$ is a normalization constant at the receiver, which does not impact detection
performance. The optimal coefficient vector $\boldsymbol{\mu}$ minimizes ${\rm mse}_n$ for a given power budget $\phi P_T$ for the information-carrying signal, i.e.,
\begin{equation}
\label{opt1}
{\rm min}_{\boldsymbol{\mu},\varsigma} \, {\rm mse}_n \qquad {\rm s.t.:}  \,{\rm Tr}\{{\bf F}_n^H {\bf F}_n\}= 1,
\end{equation}
where we use the notation ${\rm Tr}\left\{\cdot\right\}=\lim_{K \to \infty}{\rm tr}\left\{\cdot\right\}/K$. The optimal coefficient vector, $\boldsymbol{\mu}_{\rm opt}$, is provided in the following theorem.

\textit{Theorem 1}: For $K,N_T \to \infty$, the optimal coefficient vector minimizing the asymptotic average MSE of the users in the $n^{\rm th}$ cell  for the POLY precoder in (\ref{W}) is given by
\begin{equation}
\label{muopt}
\boldsymbol{\mu}_{\rm opt}=\gamma_3 \boldsymbol{\Pi}^{-1} \boldsymbol{\psi},
\end{equation}
where $\boldsymbol{\psi}=[\zeta,\,\zeta^2,\ldots,\,\zeta^{{\cal I}+1}]^T$, $[\boldsymbol{\Pi}]_{i,j}={\rm Tr}\left\{{\bf D}_{nn}\right\}\zeta^{i+j}+\left({\rm Tr}\left\{{\bf D}_{nn}
\boldsymbol{\Delta}_n\right\} + \frac{{\rm Tr}\left\{\boldsymbol{\Sigma}_n\right\}+P_{\rm AN}}{N_T p}\right) \zeta^{i+j-1}$, $\boldsymbol{\Sigma}_n=\mathbb{E}[\tilde{\bf n}_n\tilde{\bf n}^H_n]$,
$\boldsymbol{\Delta}_n={\rm diag}\left\{\frac{1+p_{\tau} \tau \sum_{m \neq n} \beta^{1}_{nm}}{1+p_{\tau} \tau \sum_{m=1}^M \beta^{1}_{nm}},\cdots, \frac{1+p_{\tau} \tau \sum_{m \neq n} \beta^K_{nm}}{1+p_{\tau}
 \tau \sum_{m=1}^M \beta^K_{nm}}\right\}$, and $P_{\rm AN}=$ \linebreak
 $q\mathbb{E}\left[{\rm Tr}\left\{{\bf G}_{nn} {\bf A}_n {\bf A}^H_n {\bf G}^H_{nn}\right\}\right]$. Furthermore, $\zeta^l$ denotes the $l^{\rm th}$-order moment of
 the sum of the eigenvalues of $\hat{\bf \overline{H}}_{nn}\hat{\bf \overline{H}}^H_{nn}$, i.e., $\zeta^l=\lim_{K \to \infty} \frac{1}{K}\sum_{k=1}^K \lambda^l_k$, which converges to $\zeta^l=\sum_{i=0}^{l-1} \binom{l}{i} \binom{l}{i+1}
 \frac{\beta^i}{l}$ for $K \to \infty$ \cite[Theorem 2]{poly}. Finally, $\gamma_3$ is chosen such that ${\rm Tr}\{{\bf F}_n^H {\bf F}_n\}= 1$ holds.
\begin{IEEEproof}
Please refer to Appendix B.
\end{IEEEproof}
We note that $\boldsymbol{\mu}_{\rm opt}$ does not depend on instantaneous channel estimates, and hence, can be computed offline.
\subsection{Computational Complexity of Data Precoding}
{We compare the computational complexity of the considered data precoders in terms of the number of floating point operations (FLOPs) \cite{flop}. Each FLOP represents one scalar complex addition or multiplication.
We assume that the coherence time of the channel is $T$ symbol intervals of which $\tau$ are used for training and $T-\tau$ are used for data transmission. Hence, the complexity required for precoding in one coherence interval
is comprised of the complexity required for generating one precoding matrix and $T-\tau$ precoded vectors. A similar complexity analysis was conducted in \cite[Section IV]{poly} for selfish data precoders without AN injection at the BS.
Since the AN injection does not affect the structure of the data precoders, we can directly adapt the results from \cite[Section IV]{poly} to the case at hand. In particular, the selfish MF, the SZF/SRCI, and the CZF/CRCI precoders require
$(2K-1)N_T (T-\tau)$,  $0.5 (K^2+K)(2 N_T-1)+K^3+K^2+K+N_T K (2K-1)+(2K-1)N_T (T-\tau)$, and $0.5 (M^2K^2+MK)(2 N_T-1)+M^3K^3+M^2 K^2+MK+N_T MK (2MK-1)+(2K-1)N_T (T-\tau)$ FLOPs per coherence interval, see  \cite[Section IV]{poly}.
In contrast, for the POLY data precoder, we obtain for the overall computational complexity $(T-\tau) \left(({\cal I}+1)(2K-1)N_T+{\cal I}(2N_T-1)K\right)$ FLOPs, which assumes implementation of the precoding operation by
Horner's rule \cite[Section IV]{poly}.}

{The above complexity expressions reveal that the additional complexity introduced by collaborative data precoders compared to selfish data precoders is at most a factor of $M^3$. In addition, the complexity savings achieved with the
POLY data precoder compared to the SZF/SRCI data precoders increase with increasing $K$ for a given $T$. We note however that, regardless of their complexity, POLY data precoders are attractive as they avoid the stability issues
that may arise in fixed point implementation of large matrix inverses.}
\section{Linear AN Precoders for Secure Massive MIMO}\label{s4}
In this section, we investigate the performance of selfish and collaborative NS (S/CNS) and random AN precoders. In addition, a novel POLY AN precoder is derived. {To the best of the authors' knowledge, POLY AN precoding has not been
considered in the literature before. }
\subsection{Analysis of Existing AN Precoders}\label{s4a}
For a given dimensionality of the AN precoder, $L$, the secrecy rate depends on the AN precoder only via the AN leakage, $Q$, given in (\ref{eq8}), which affects the SINR of the MT. Furthermore, the optimal POLY data
precoder coefficients in (\ref{muopt}) are affected by the AN precoder via the leakage term $P_{\rm AN}$. In this subsection, for $N_T\to\infty$, we will provide closed-form expressions for $Q$ and $P_{\rm AN}$
for the SNS, CNS, and random AN precoders.
\subsubsection{SNS AN Precoder}
The SNS AN precoder of the $n^{\rm th}$ BS is given by \cite{negi}
\begin{equation}
\label{sn}
{\bf A}_n={\bf I}_{N_T}-\hat{\bf H}^H_{nn} \left(\hat{\bf H}_{nn} \hat{\bf H}^H_{nn}\right)^{-1} \hat{\bf H}_{nn},
\end{equation}
which has rank $L=N_T-K$ and exists only if $\beta<1$. We divide the corresponding AN leakage $Q_{\rm SNS}$ into an inter-cell AN leakage $Q_o^{\rm SNS}$ and an intra-cell AN leakage $Q_i^{\rm SNS}$, where
$Q_{\rm SNS}=Q_o^{\rm SNS}+Q_i^{\rm SNS}$. For the SNS AN precoder, $Q_o^{\rm SNS}$ is obtained as
\begin{equation}
Q_o^{\rm SNS}=\sum_{m \neq n} \beta^k_{mn} \mathbb{E}\bigg[{\bf h}^k_{mn} {\bf A}_{m} {\bf A}^H_m ({\bf h}^k_{mn})^H\bigg]=\mathbb{E}\bigg[{\rm tr}\left\{{\bf A}_{m} {\bf A}^H_m\right\}\bigg]\sum_{m \neq n}^M \beta^k_{mn}= (N_T-K)\sum_{m \neq n}^M \beta^k_{mn},
\end{equation}
where we exploited \cite[Lemma 11]{muller} and the independence of ${\bf A}_{m}$ and ${\bf h}^k_{mn}$. In contrast, the intra-cell AN leakage power is given by $Q_i^{\rm SNS}=$
\begin{equation}\nonumber
\beta^k_{nn} \mathbb{E}\bigg[{\bf h}^k_{nn} {\bf A}_{n} {\bf A}^H_n ({\bf h}^k_{nn})^H\bigg]=\beta^k_{nn}\mathbb{E}\bigg[\tilde{\bf h}^k_{nn} {\bf A}_{n} {\bf A}^H_n (\tilde{\bf h}^k_{nn})^H\bigg]=(N_T-K)\beta^k_{nn}\frac{1+
p_\tau \tau \sum_{m \neq n}^M \beta^k_{nm}}{1+p_\tau \tau \sum_{m=1}^M \beta^k_{nm}},
\end{equation}
as the SNS AN  precoder matrix lies in the null space of the estimated channels of all $K$ MTs in the $n^{\rm th}$ cell. Similarly, the AN leakage relevant for computation of the POLY data precoder is obtained as
\begin{equation}\label{aaa}
P_{\rm AN}^{\rm SNS}=(1-\phi)P_T \lim_{K\to\infty} \frac{1}{K} \sum_{k=1}^K \beta^k_{nn} \frac{1+p_\tau \tau \sum_{m \neq n}^M \beta^k_{nm}}{1+p_\tau \tau \sum_{m=1}^M \beta^k_{nm}}.
\end{equation}
\subsubsection{CNS AN Precoder}
For the CNS AN precoder at the $n^{\rm th}$ BS, the AN is designed to lie in the null space of the estimated channels between all $MK$ MTs and the BS, i.e.,
\begin{equation}
{\bf A}_n={\bf I}_{N_T}-\hat{\bf H}^H_n \left(\hat{\bf H}_n \hat{\bf H}^H_n\right)^{-1} \hat{\bf H}_n,
\end{equation}
which has rank $L=N_T-MK$ and exists only if $\beta < 1/M$. The corresponding AN leakage to the $k^{\rm th}$ MT in the $n^{\rm th}$ cell is given by
\begin{equation}
Q_{\rm CNS}=\sum_{m=1}^M \beta^k_{mn} \mathbb{E}\bigg[{\bf h}^k_{mn} {\bf A}_{m} {\bf A}^H_m ({\bf h}^k_{mn})^H\bigg]= (N_T-MK) \sum_{m=1}^M \beta^k_{mn}\frac{1+p_\tau
\tau \sum_{l \neq m}^M \beta^k_{ml}}{1+p_\tau \tau \sum_{l=1}^M \beta^k_{ml}}.
\end{equation}
Furthermore, the CNS AN precoder results in the same $P_{\rm AN}$ as the SNS AN precoder, cf.~(\ref{aaa}).
\subsubsection{Random AN Precoder}
For the random precoder, all elements of ${\bf A}_n$ are i.i.d.~random variables independent of the channel \cite{zhu}, i.e., ${\bf A}_n$ has rank $L=N_T$.
Hence, ${\bf h}^k_{mn}$ and ${\bf A}_{m}$, $\forall m$, are mutually independent, and we obtain
\begin{equation}
Q_{\rm random}=\sum_{m=1}^M \beta^k_{mn} \mathbb{E}\bigg[{\bf h}^k_{mn} {\bf A}_{m} {\bf A}^H_m ({\bf h}^k_{mn})^H\bigg]=N_T\sum_{m=1}^M \beta^k_{mn}.
\end{equation}
Furthermore, we obtain $P_{\rm AN}^{\rm random}=(1-\phi)P_T \lim_{K\to\infty} \frac{1}{K} \sum_{k=1}^K \beta^k_{nn}$.

\textit{Remark 2:} If the power and time allocated to channel estimation are very small, i.e., $\tau p_\tau \to 0$, the S/CNS AN precoders yield the same $qQ$ and $P_{\rm AN}$ as the random AN precoder. This suggests
that in this regime all considered AN precoders achieve a similar SINR performance for a given MT. However, for $\tau p_\tau >0$, the S/CNS AN precoders cause less AN leakage resulting in an improved SINR performance compared to the
random precoder at the expense of a higher complexity.
\subsection{POLY AN Precoder}\label{s4b}
To mitigate the high computational complexity imposed by the matrix inversion required for the S/CNS AN precoders, while achieving an improved performance compared to the random AN precoder, we propose a
POLY AN precoder. Similar to the POLY data precoder, we concentrate on the selfish design because of the desired low complexity, and hence, set $L=N_T-K$. The proposed POLY AN precoder is given by
\begin{equation}
\label{V}
{\bf A}_n={\bf I}_{N_T}-\hat{\bf \overline{H}}^H_{nn} \left(\sum_{i=0}^{\cal J} \nu_j \left(\hat{\bf \overline{H}}_{nn} \hat{\bf \overline{H}}^H_{nn} \right)^j \right)\hat{\bf \overline{H}}_{nn},
\end{equation}
where $\boldsymbol{\nu}=[\nu_0,\ldots,\nu_{\cal J}]^T$ contains the real-valued coefficients of the AN precoder polynomial, which have to be optimized. In particular, $\boldsymbol{\nu}$ is optimized for minimization of the
asymptotic average AN leakage caused to all MTs in the $n^{\rm th}$ cell $P_{\rm AN}$. The corresponding optimization problem is formulated as
\begin{equation}
\label{opt2}
{\rm min}_{\boldsymbol{\nu}} \,P_{\rm AN}=q\mathbb{E}\bigg[{\rm Tr}\{{\bf G}_{nn} {\bf A}_n{\bf A}_n^H   {\bf G}_{nn}^H\}\bigg]\quad {\rm s.t.:} {\rm Tr}\{{\bf A}_n^H {\bf A}_n\}= 1/\beta-1.
\end{equation}
The solution of (\ref{opt2}) is provided in the following theorem.

\textit{Theorem 2}: For $K,N_T \to \infty$, the optimal coefficient vector minimizing the asymptotic average AN leakage caused to the users in the $n^{\rm th}$ cell for the AN precoder structure in
(\ref{V}) is given by
\begin{equation}
\boldsymbol{\nu}_{\rm opt}=\boldsymbol{\Sigma}^{-1} \boldsymbol{\omega},
\end{equation}
where $[\boldsymbol{\Sigma}]_{i,j}=\zeta^{i+j+1}+\epsilon \zeta^{i+j}$ and $\boldsymbol{\omega}=[\zeta^2+\epsilon \zeta,\ldots,\zeta^{{\cal J}+2}+\epsilon \zeta^{{\cal J}+1}]$. Here, $\zeta^l$ denotes again the
$l^{\rm th}$ order moment of the sum of the eigenvalues of matrix $\hat{\bf \overline{H}}_{nn}\hat{\bf \overline{H}}^H_{nn}$, cf.~Theorem 1. $\epsilon$ is chosen such that ${\rm Tr}\{{\bf A}_n^H {\bf A}_n\}= 1/\beta-1$.
\begin{IEEEproof}
Please refer to Appendix C.
\end{IEEEproof}
\subsection{Computational Complexity of AN Precoding}
{Similarly to the data precoders, the complexity of the AN precoders is evaluated in terms of the number of flops required per coherence interval $T$. For the SNS AN precoder, the computation of ${\bf A}_n$
in (\ref{sn}) requires the computation and inversion of a $K \times K$ positive definite matrix, which entails $0.5(K^2+K) (2N_T-1)+K^3+K^2+K$ FLOPs \cite{flop}, and the multiplication of an $N_T \times K$, an $K \times K$, and an
$K \times N_T$ matrix, which entails $N_T(N_T+K)(2K-1)$ FLOPs \cite{flop}. Furthermore, the $T-\tau$ vector-matrix multiplications required for AN precoding entail a complexity of $(2N_T-1)N_T$ FLOPs \cite{flop}, respectively.
Hence, the overall complexity is $0.5(K^2+K) (2N_T-1)+K^3+K^2+K+N_T(N_T+K)(2K-1)+(2N_T-1)N_T(T-\tau)$ FLOPs. Similarly, for the CNS AN precoder, we obtain a complexity of $0.5((MK)^2+MK) (2N_T-1)+(MK)^3+(MK)^2+
MK+N_T(N_T+MK)(2MK-1)+(2N_T-1)N_T(T-\tau)$ FLOPs, whereas the random AN precoder entails a complexity of $(2N_T-1)N_T(T-\tau)$ FLOPs as only the AN vector-matrix multiplications are required.}

{Similar to the precoded data vector \cite[Section IV]{poly}, the POLY precoded AN vector can be generated using Horner's rule. Hence, based on (\ref{V}), the transmitted AN vector in the $n^{\rm th}$ cell can be obtained as
\begin{equation}\label{Ahorner}
{\bf A}_n {\bf z}_n={\bf z}_n-\left(\nu_0 \hat{\bf \overline{H}}^H_{nn}\hat{\bf \overline{H}}_{nn} \left({\bf z}_n+\frac{\nu_1}{\nu_0} \hat{\bf \overline{H}}^H_{nn}\hat{\bf \overline{H}}_{nn} \left({\bf z}_n+\ldots\right)\right)\right).
\end{equation}
Hence, ${\bf A}_n {\bf z}_n$ can be computed efficiently by first multiplying $\hat{\bf \overline{H}}_{nn}$ with ${\bf z}_n$, which requires $(2N_T-1)K$ FLOPs, then multiplying $\hat{\bf \overline{H}}^H_{nn}$ with the resulting vector, which requires
$(2K-1)N_T$ FLOPs, adding ${\bf z}_n$ to the newly resulting vector, and repeating similar operations $({\cal J}+1)$ times, see  \cite{polypio,poly} for details of Horner's rule. Overall, this leads to a complexity of
$({\cal J}+1)\left((2K-1)N_T+(2N_T-1)K\right)(T-\tau)$ FLOPs.}
\section{Comparison of Linear Data and AN Precoders}\label{s5}
In this subsection, we compare the secrecy performances of the considered data and AN precoders. Thereby, in order to get tractable results, we focus on the relative performances of SZF, CZF, and MF \cite{zhu} data precoders
and SNS, CNS, and random AN precoders. The performances of SRCI, CRCI, and POLY data precoders and the POLY AN precoder will be investigated via numerical and simulation results in Section \ref{s6}.

In order to gain some insight for system design and analysis, we adopt a simplified path-loss model. In particular, we assume the path losses are given by
\begin{equation}
\beta^k_{mn}=\begin{cases} 1,& m=n\\ \rho, & {\rm otherwise}\end{cases}
\label{eq21}
\end{equation}
where $\rho \in [0,1]$ denotes the inter-cell interference factor. For this simplified model, $a$ and $c$ in (\ref{Cup}) simplify to $a=1+(M-1)\rho$ and $c=1+(M-1)\rho^2$. Furthermore, the SINR expressions of the linear data precoders
considered in Section \ref{s3a} and the MF precoder considered in \cite{zhu} can be simplified considerably and are provided in Table \ref{tablea}, where we use the normalized AN leakage
$\tilde Q = Q/L$. The expressions for the normalized AN leakage $\tilde Q$, the asymptotic average AN leakage $P_{\rm AN}$, and the dimensionality $L$ of the considered linear AN precoders
are given in Table \ref{tableb}.
\begin{table*}
\centering
\caption{SINR of the $k^{\rm th}$ MT in the $n^{\rm th}$ cell for linear data precoding and the simplified path-loss model in (\ref{eq21}). For this model, $\hat{\Gamma}_{\rm SRCI}$ and $\hat{\Gamma}_{\rm CRCI}$ simplify to
$\hat{\Gamma}_{\rm SRCI}=\frac{\Gamma_{\rm SRCI} \theta}{\Gamma_{\rm SRCI} \vartheta+1}$ and $\hat{\Gamma}_{\rm CRCI}=\frac{\Gamma_{\rm CRCI} \theta}{\Gamma_{\rm CRCI} \vartheta+1}$
where $\Gamma_{\rm SRCI}=\frac{\beta\phi}{\beta \phi \rho (M-1)+(1-\phi)\beta \tilde{Q}+\frac{\beta}{P_T}}$, $\Gamma_{\rm CRCI}=\frac{\beta\phi}{(1-\phi)\beta \tilde{Q}+\frac{\beta}{P_T}}$, $\theta=\frac{p_{\tau} \tau}{1+a p_{\tau}\tau}$,
and $\vartheta=\frac{1+(M-1)\rho p_{\tau}\tau}{1+a p_{\tau}\tau}$.}
\begin{tabular}{|c|c|}
  \hline
Data Precoder   & $\gamma_{nk}$ \\
  \hline
  SZF & $\frac{\theta \phi (1-\beta)}{(1-\phi)\beta \tilde{Q} +\beta\phi (a-\theta)+(M-1)\rho^2 \theta \phi (1-\beta)+\beta/P_T}$\\
  \hline
  SRCI & $\frac{1}{1/\mathcal{G}(\beta,\beta/\hat{\Gamma}_{\rm SRCI})+(M-1)\rho}$ \\
  \hline
  CZF & $\frac{\theta\phi  (1-M \beta)}{(1-\phi)\beta \tilde{Q}+\beta \phi a(1-\theta)+(M-1)\rho^2 \theta \phi (1-M \beta)+\beta/P_T}$\\\hline
  CRCI & $\frac{1}{1/\mathcal{G}(a \beta,a \beta/\hat{\Gamma}_{\rm CRCI})+(M-1)\rho}$\\
  \hline
  MF &  $\frac{\theta\phi}{(1-\phi)\beta \tilde{Q}+\beta\phi a+(M-1)\rho^2 \theta \phi+\beta/P_T}$\\
  \hline
\end{tabular}\label{tablea}
\end{table*}

\begin{table*}
\centering
\caption{AN leakage for simplified path-loss model in (\ref{eq21}). $\theta$ and $\vartheta$ are defined in the caption of Table \ref{tablea}. }
\begin{tabular}{|c|c|c|c|}
  \hline
AN Precoder    & $\tilde{Q}$ & $P_{\rm AN}$ & $L$\\
  \hline
 SNS & $(a-\theta)$& $ (1-\phi)P_T \vartheta$ & $N_T-K$\\
  \hline
  CNS & $a(1-\theta)$& $(1-\phi)P_T \vartheta$ & $N_T-MK$\\
  \hline
  Random & $a$& $(1-\phi)P_T$ & $N_T$\\
  \hline
\end{tabular}\label{tableb}
\end{table*}
\subsection{Comparison of SZF, CZF, and MF Data Precoders}\label{s5a}
In this subsection, we compare the performances achieved with SZF, CZF, and MF data precoders for a given AN precoder, i.e., $L$ and $\tilde Q$ are fixed. Since the upper bound on the capacity of the eavesdropper channel is independent of the
adopted data precoder, cf.~Section \ref{s2c}, we compare the considered data precoders based on their SINRs. Exploiting the results in Table \ref{tablea}, we obtain the following relations between
$\gamma^{\rm SZF}_{nk}$, $\gamma^{\rm CZF}_{nk}$, and $\gamma^{\rm MF}_{nk}$:
\begin{equation}
\label{eq25}
\frac{\gamma^{\rm SZF}_{nk}}{\gamma^{\rm MF}_{nk}}=1+\beta (c\gamma^{\rm SZF}_{nk}-1)\qquad {\rm and}\qquad \frac{\gamma^{\rm CZF}_{nk}}{\gamma^{\rm SZF}_{nk}}=
\frac{1-M \beta}{1-\beta}+ \frac{a(a-1)\beta }{1-\beta} \gamma^{\rm CZF}_{nk}.
\end{equation}
Hence, for $\gamma^{\rm SZF}_{nk}>\gamma^{\rm MF}_{nk}$, we require $\gamma^{\rm SZF}_{nk}>1/c=1/(1+\rho^2(M-1))$, and for $\gamma^{\rm CZF}_{nk}>\gamma^{\rm SZF}_{nk}$, we need
$\gamma^{\rm CZF}_{nk}>1/(\rho a)=1/[\rho(1+\rho(M-1))]$. As expected, (\ref{eq25}) suggests that for a lightly loaded system, i.e., $\beta\to 0$, all three precoders have a similar performance, i.e.,
$\gamma^{\rm CZF}_{nk} \approx \gamma^{\rm SZF}_{nk} \approx \gamma^{\rm MF}_{nk}$. In the following, we investigate the impact of the number of MTs and the pilot
power on the relative performances of the considered data precoders.

1) \textit{Number of MTs}: From (\ref{eq25}), we find that for $\gamma^{\rm SZF}_{nk}>\gamma^{\rm MF}_{nk}$ and $\gamma^{\rm CZF}_{nk}>\gamma^{\rm SZF}_{nk}$ to hold, the number of MTs has to meet $K<K_{{\rm SZF} > {\rm MF}}$ and $K<K_{{\rm CZF} > {\rm SZF}}$, where $K_{{\rm SZF} > {\rm MF}} =$
\begin{equation}
\label{betazf}
 \frac{\theta \phi N_T}{(1-\phi)\tilde{Q}+a\phi+1/P_T}\quad {\rm and} \quad K_{{\rm CZF} > {\rm SZF}}=\frac{\rho \phi \theta N_T}{(1-\phi)\tilde{Q}+[a(1-\theta)+\rho \theta M]\phi+1/P_T},
\end{equation}
respectively. Interestingly, both the maximum numbers of MTs for which the SZF data precoder is advantageous compared to the MF data precoder, $K_{{\rm SZF} > {\rm MF}} $, and the maximum number of MTs for which
the CZF data precoder is advantageous compared to the SZF data precoder, $K_{{\rm CZF} > {\rm SZF}}$, decrease with increasing AN leakage, $\tilde Q$, and increasing number of cells, $M$, but increase with the amount
of resources dedicated to channel estimation, $p_\tau\tau$ (via $\theta$), and consequently with the channel estimation quality. However, while $K_{{\rm SZF} > {\rm MF}} $ decreases with increasing inter-cell interference factor, $\rho$ (via $a$),
$K_{{\rm CZF} > {\rm SZF}}$ increases.

2) \textit{Pilot Energy}: From (\ref{eq25}), we find that for $\gamma^{\rm SZF}_{nk}>\gamma^{\rm MF}_{nk}$ and $\gamma^{\rm CZF}_{nk}>\gamma^{\rm SZF}_{nk}$ to hold, pilot energy
$p_\tau \tau$ has to fulfill
\begin{equation}
\label{przf}
p_\tau \tau>(p_\tau \tau)_{{\rm SZF} > {\rm MF}} = \frac{1}{\frac{\phi (1-\beta)/\beta+1}{a+1/P_T}-a} \quad {\rm and}\quad p_\tau \tau>(p_\tau \tau)_{{\rm CZF} > {\rm SZF}}=\frac{1}{\frac{\rho \phi (1-\beta)/\beta+1}{a+1/P_T}-a},
\end{equation}
where we have assumed that SNS AN precoding is adopted, i.e., $\tilde{Q}=a-\theta$, to arrive at insightful expressions. Similar results can be obtained for other AN precoders.
From (\ref{przf}), we observe that MF, SZF, and CZF data precoding are preferable if
$0<p_\tau \tau <(p_\tau \tau)_{{\rm SZF} > {\rm MF}}$, $(p_\tau \tau)_{{\rm SZF} > {\rm MF}}\le p_\tau \tau< (p_\tau \tau)_{{\rm CZF} > {\rm SZF}}$, and $p_\tau \tau\ge (p_\tau \tau)_{{\rm CZF} > {\rm SZF}}$,
respectively. In general, the more MTs are in the system (i.e., the larger $\beta$), the larger the pilot energy has to be to make SZF and CZF data precoding beneficial. In fact, from (\ref{przf}) we observe that
if $\beta$ exceeds $\beta_{\rm MF}=\phi/[a^2+a/P_T+\phi-1]$, MF data precoding is always preferable regardless of the value of $p_\tau\tau$. Similarly, if $\beta$ exceeds $\beta_{\rm SZF}=
\phi\rho/[a^2+a/P_T+\phi\rho-1]$, SZF data precoding is always preferable compared to CZF data precoding regardless of the value of $p_\tau\tau$.
\subsection{Comparison of SNS, CNS, and MF AN Precoding}\label{s5b}
In this subsection, we analyze the impact of the AN precoders on the secrecy rate. AN precoders affect the ergodic capacity of the eavesdropper via $L$ and the achievable rate of the MT via the leakage, $\tilde Q$. Since the upper bound
on the ergodic secrecy rate of the eavesdropper in (\ref{Cup}) is a decreasing function in $L$, we have
\begin{equation}
 C_{nk}^{\rm eve}|_{\rm random}\le  C_{nk}^{\rm eve}|_{\rm SNS} \le  C_{nk}^{\rm eve}|_{\rm CNS}.
\label{a-1}
\end{equation}
On the other hand, from Table \ref{tableb}, we observe $\tilde{Q}_{\rm random} \ge \tilde{Q}_{\rm SNS} \ge \tilde{Q}_{\rm CNS}$. Since according to Table \ref{tablea} the SINRs for all data precoders are decreasing functions of $\tilde{Q}$,
for a given data precoder, we obtain for the lower bound on the ergodic rate of the $k^{\rm th}$ MT in the $n^{\rm th}$ cell
\begin{equation}
 R_{nk}|_{\rm random}\le  R_{nk}|_{\rm SNS} \le  R_{nk}|_{\rm CNS}.
\label{a-2}
\end{equation}
Considering (\ref{a-1}), (\ref{a-2}), and the expression for the ergodic secrecy rate, $R_{nk}^{\rm sec}=[R_{nk}-C^{\rm eve}_{nk}]^+$, it is not a priori clear which AN precoder has the best performance. In fact, our numerical results
in Section \ref{s6} confirm that it depends on the system parameters (e.g.~$\alpha$, $\beta$, $M$, $p_\tau\tau$, and $\rho$) which AN precoder is preferable.
\subsection{Ergodic Secrecy Rate Analysis}\label{s5c}
In this subsection, we provide closed-form results for the ergodic secrecy rate for SZF, CZF, and MF data precoding for the simplified path-loss model in (\ref{eq21}). Thereby, the simplified path-loss model is extended also to the
eavesdropper, i.e., $\beta_{nE}=1$ and $\beta_{mE}=\rho$, $m\ne n$, is assumed.

Combining (\ref{secnk}), (\ref{Cup}), and the results in Table \ref{tablea}, we obtain the following lower bounds for the ergodic secrecy rate of the $k^{\rm th}$ MT in the $n^{\rm th}$ cell:
\begin{equation}
\label{rmfzf}
R^{{\rm sec}}_{nk} \geq \begin{cases}\bigg[\log_2 \left(\frac{(\tilde{Q}+1/P_T)\beta+(a-\tilde{Q})\beta \phi+c\theta\phi}{(\tilde{Q}+1/P_T)\beta+(a-\tilde{Q})\beta \phi+(c-1)\theta\phi}\cdot \frac{-\chi \phi+\chi}{(1-\chi)\phi+\chi}\right)\bigg]^+& \quad {\rm for~MF},\\\bigg[\log_2 \left(\frac{(\tilde{Q}+1/P_T)\beta+(a-\theta-\tilde{Q})\beta \phi+c\theta(1-\beta)\phi}{(\tilde{Q}+1/P_T)\beta+(a-\theta-\tilde{Q})\beta \phi+(c-1)\theta(1-\beta)\phi}\cdot \frac{-\chi \phi+\chi}{(1-\chi)\phi+\chi}\right)\bigg]^+& \quad {\rm for~SZF},\\\bigg[\log_2 \left(\frac{(\tilde{Q}+1/P_T)\beta+(a-a\theta-\tilde{Q})\beta \phi+c\theta(1-M\beta)\phi}{(\tilde{Q}+1/P_T)\beta+(a-a\theta-\tilde{Q})\beta \phi+(c-1)\theta(1-M\beta)\phi}\cdot \frac{-\chi \phi+\chi}{(1-\chi)\phi+\chi}\right)\bigg]^+& \quad {\rm for~CZF},\end{cases}
\end{equation}
where $\chi=\frac{a \beta}{\alpha}-\frac{\beta c N_T}{a L}$, and $\tilde{Q}$ and $L$ are given in Table \ref{tableb} for the considered AN precoders.  Eq.~(\ref{rmfzf}) is easy to evaluate and reveals how the ergodic secrecy rate of the three considered
data precoders depends on the various system parameters. To gain more insight, we determine the maximum value of $\alpha$ which admits a non-zero secrecy rate. This value is denoted by $\alpha_s$ in the following, and can be shown to be a
decreasing function of $\phi$ for all conidered data precoders. Hence, we find $\alpha_s$ by setting $R^{\rm sec}_{nk}=0$ in (\ref{rmfzf}) and letting $\phi \to 0$. This leads to
\begin{equation}
\label{alphas}
\alpha_s=\begin{cases}
\frac{a^2 \theta}{\tilde{Q}a+c\theta N_T/L+a/P_T} & {\rm for~MF},\\
\frac{(1-\beta)a^2 \theta}{\tilde{Q}a+c\theta (1-\beta)N_T/L+a/P_T} & {\rm for~SZF},\\
 \frac{(1-M\beta)a^2 \theta}{\tilde{Q}a+c\theta (1-M \beta)N_T/L+a/P_T} & {\rm for~CZF}.
\end{cases}
\end{equation}
Eq.~(\ref{alphas}) reveals that for a given AN precoder, independent of the system parameters, the MF data precoder can always tolerate a larger number of eavesdropper antennas than the SZF data precoder, which in turn can always tolerate a
larger number of eavesdropper antennas than the CZF data precoder. This can be explained by the fact that the high AN transmit power required to combat a large number of eavesdropper antennas drives the receiver of the desired MT into the
noise-limited regime, where the MF data precoder has a superior performance compared to the S/CZF data precoders. On the other hand, since $\alpha_s$ depends on both $\tilde{Q}$ and $L$, it is not a priori clear which AN precoder can tolerate
the largest number of eavesdropper antennas. For a lightly loaded network with small $\beta$ and small $M$, according to Table \ref{tableb}, we have $L\approx N_T$ for all three AN precoders. Hence, in this case, we expect the CNS AN precoder to
outperform the SNS and random AN precoders as it achieves a smaller $\tilde Q$. On the other hand, for a heavily loaded network with large $\beta$ and $M$, the value of $\alpha_s$ of the CNS AN precoder is compromised by its small value of $L$
and SNS and even random AN precoders are expected to achieve a larger $\alpha_s$.
\section{Performance Evaluation}\label{s6}
In this section, we evaluate the performance of the considered secure multi-cell massive MIMO system. We consider cellular systems with $M=2$ and $M=7$ hexagonal cells, respectively, and to gain insight for system design, we adopt the simplified path-loss model introduced in Section \ref{s5}, i.e., the severeness of the inter-cell interference is only characterized by the parameter $\rho \in (0,1]$. The pilot sequence length is $\tau=K$. The simulation results for the ergodic secrecy rate of the $k^{\rm th}$ MT in the $n^{\rm th}$ cell are based on (\ref{secnk}), (\ref{Ceve}), and the expression for the ergodic rate of the MT \cite[Eq.~(8)]{zhu} and are averaged over $5,000$ random channel realizations. {Note that, in this paper, we consider the ergodic secrecy rate of a certain MT, i.e., the $k^{\rm th}$ MT in the $n^{\rm th}$ cell. The cell sum secrecy rate can be obtained by multiplying the secrecy rate of the $k^{\rm th}$ MT by the number of MTs, $K$, as for the considered channel model, all MTs in the $n^{\rm th}$ cell achieve the same secrecy rate.} The values of all relevant system parameters are provided in the captions of the figures. To enable a fair comparison, throughout this section, we adopted the selfish SNS AN precoder when we compare different data precoders
and the selfish ZF data precoder when we compare different AN precoders.
\subsection{Ergodic Capacity of the Eavesdropper for Conventional AN Precoders} \label{s60}
\begin{figure}
  \centering
    \includegraphics[width=3.7in]{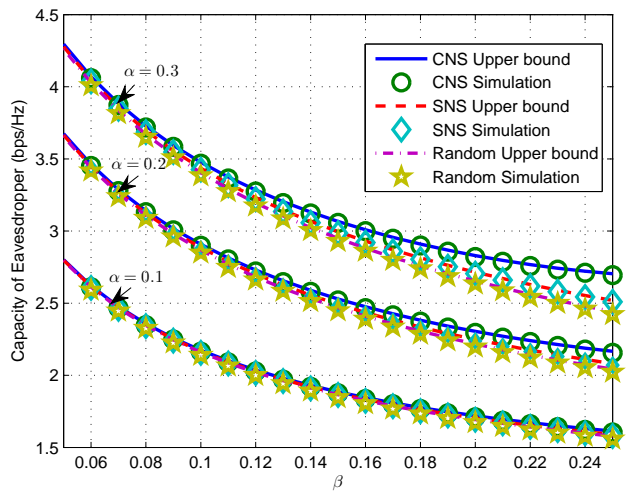}\\
    \caption{Ergodic capacity of the eavesdropper vs.~the normalized number of MTs in the cell, $\beta$, for a system with $N_T=200$, $\phi=0.75$, $P_T=10$ dB, $\rho=0.3$, and $M=2$.}\label{Fig0}
  \end{figure}

In Fig.~\ref{Fig0}, we show the ergodic capacity of the eavesdropper for the considered conventional AN precoders. First, we note that the upper bound in  (\ref{Cup}) is very tight {since the number of BS antennas is large
($N_T=200$) and $\alpha<a^2L/(cN_T)$ holds for all considered AN precoders and all consider values of $\alpha$ and $\beta$.}
Furthermore, as $\beta$ increases, the ergodic capacity of all AN precoders decreases since the power allocated to the information-carrying signal of the user that the eavesdropper tries to intercept decreases with increasing $\beta$ as the total
power allocated to the information-carrying signals of all users is fixed. As expected, the eavesdropper's capacity benefits from larger values of $\alpha$. Furthermore, as predicted in (\ref{a-1}), because of their different values of $L$, the CNS AN
precoder yields the largest eavesdropper capacity, while the random AN precoder yields the lowest. The performance differences between the different AN precoders diminish for small values of $\alpha$ and $\beta$ as the dependence of the
eavesdropper capacity on $L$ becomes negligible for small $\alpha$, cf.~(\ref{Cup}), and $L\approx N_T$ holds for all precoders for small $\beta$, cf.~Table \ref{tableb}.

\subsection{Ergodic Secrecy Rate for Conventional Linear Data Precoders} \label{s6a}
\begin{figure}[t]
\centering
 \begin{minipage}[b]{0.45\linewidth} \hspace*{-1cm}
  \centering
  \includegraphics[width=3.7in]{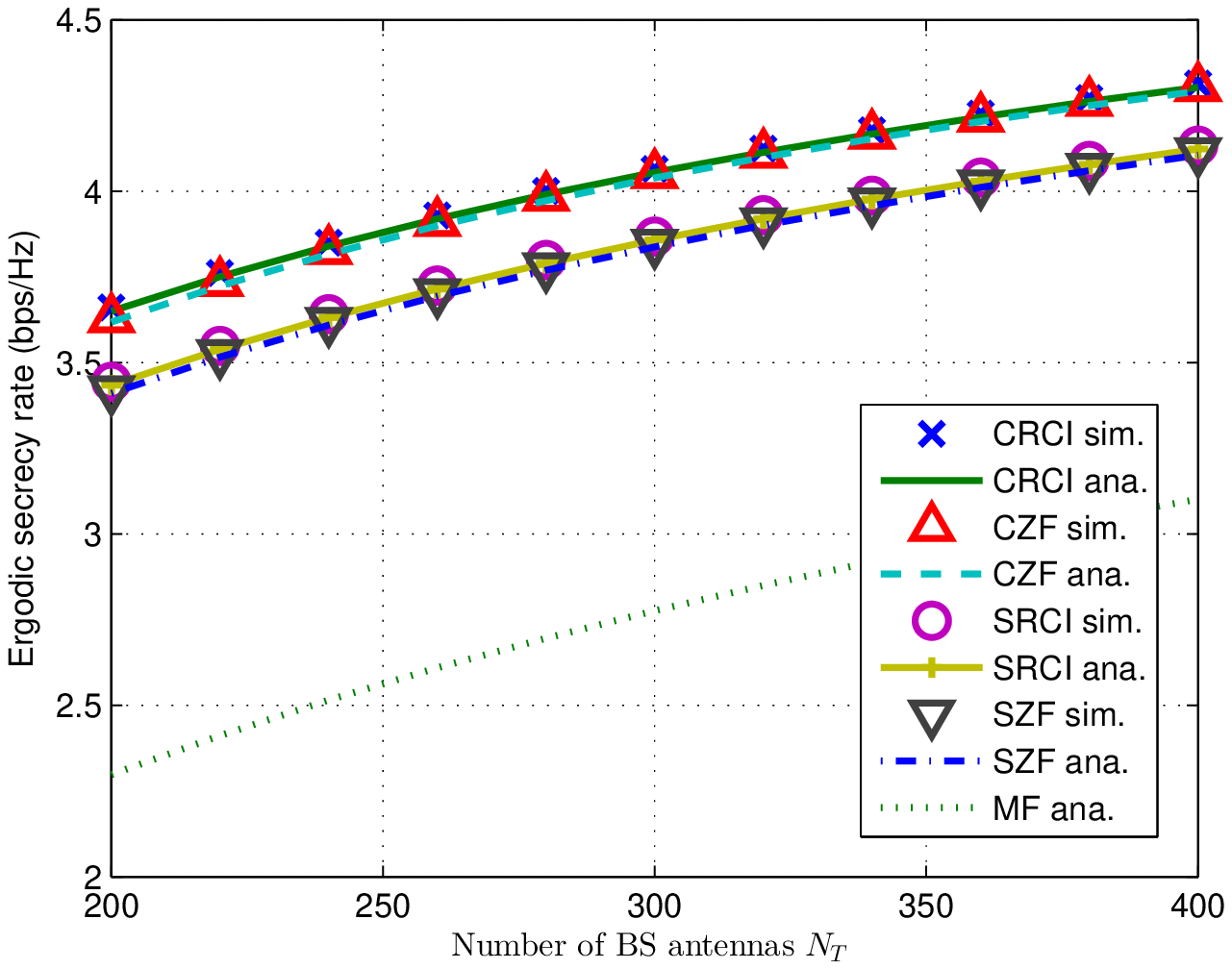}\vspace*{-8mm}\\
  \caption{Analytical and simulation results for the ergodic secrecy rate vs.~the number of BS antennas, $N_T$, for a lightly loaded network with $\phi=0.75$, $P_T=10$ dB, $p_\tau=P_T/K$, $\alpha=0.1$, $K=10$, $\rho=0.1$, and $M=2$.}
  \label{Fig1}
 \end{minipage}\hspace*{1.1cm}
 \begin{minipage}[b]{0.45\linewidth} \hspace*{-1cm}
  \centering
  \includegraphics[width=3.7in]{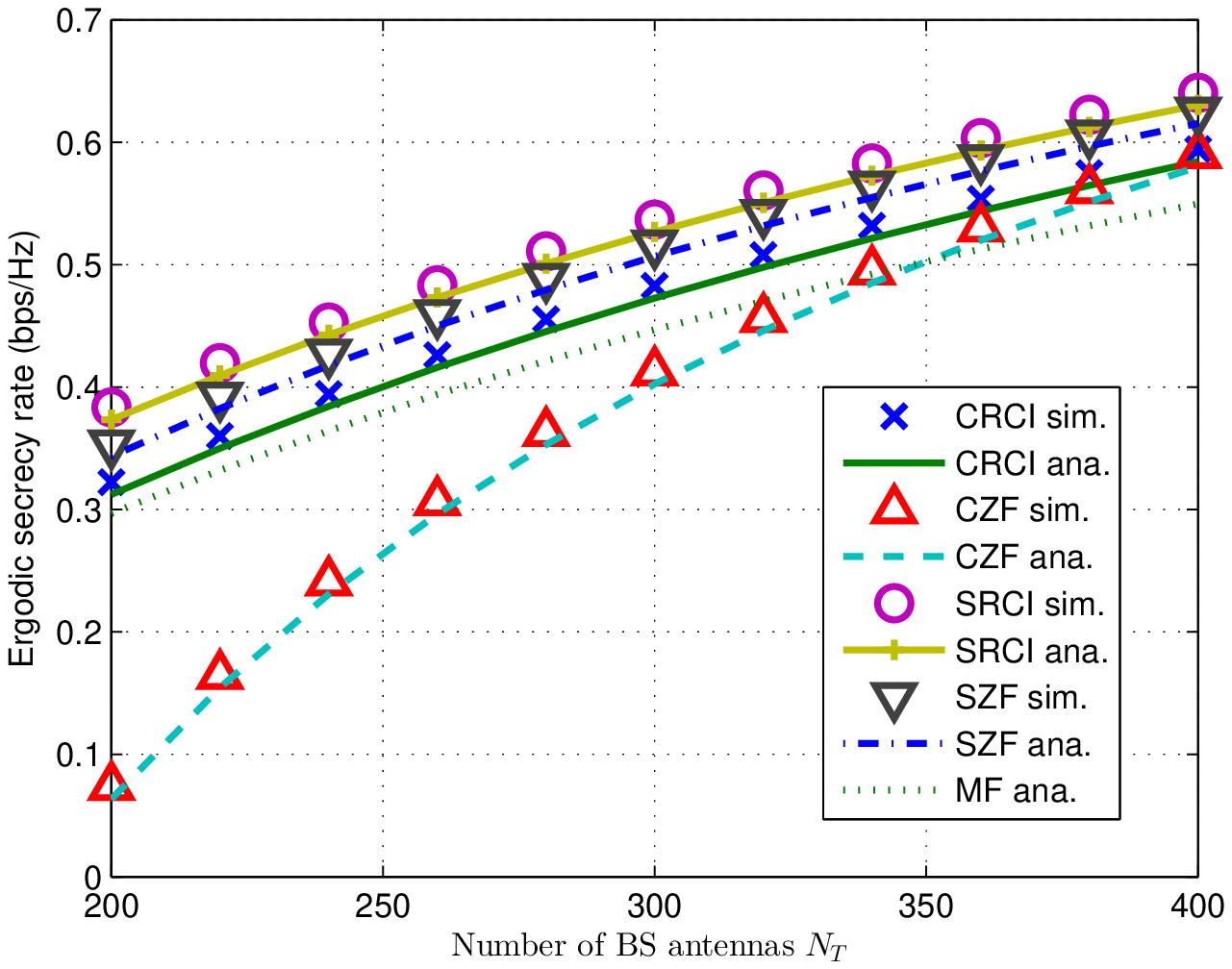}\vspace*{-8mm}\\
  \caption{Analytical and simulation results for the ergodic secrecy rate vs.~the number of BS antennas, $N_T$, for a dense network with $\phi=0.75$, $P_T=10$ dB, $p_\tau=P_T/K$, $\alpha=0.1$, $K=20$, $\rho=0.3$, and $M=7$.}
  \label{Fig2}
 \end{minipage}\hspace*{1.1cm}
  \end{figure}

In Figs.~\ref{Fig1} and \ref{Fig2}, we show the ergodic secrecy rates of the $k^{\rm th}$ MT in the $n^{\rm th}$ cell vs.~the number of BS antennas for the MF, SZF, CZF, SRCI, and CRCI data precoders for
a lightly loaded and a dense network, respectively, and a fixed power allocation factor of $\phi=0.75$. In both figures, the analytical results were obtained from (\ref{secnk}), (\ref{Ceve}), and (\ref{srci-opt}) for the SRCI data precoder,  (\ref{crci-opt}) for
the CRCI data precoder, and (\ref{rmfzf}) for the MF, SZF, and CZF data precoders. For all considered precoders, the analytical results provide a tight lower bound for
the ergodic secrecy rates obtained by simulations. Furthermore, as expected, the RCI data precoders outperform the ZF data precoders for both the selfish and the collaborative strategies, but the performance gap
diminishes with increasing number of BS antennas.

For the lightly loaded network in Fig.~\ref{Fig1}, we assume $M=2$ cells, $K=10$ users, and a small inter-cell interference factor of $\rho=0.1$. For this scenario, the collaborative designs outperform the selfish designs and
C/SZF precoding yields a large performance gain compared to MF precoding. This is expected from our analysis in Section \ref{s5a} as for the parameters valid for Fig.~\ref{Fig1}, we obtain from (\ref{betazf}),
$K_{{\rm SZF} > {\rm MF}}  \approx 250$ and $K_{{\rm CZF} > {\rm SZF}} \approx 60$ for $N_T=400$. Intuitively, as the network is only lightly loaded, the collaborative data precoder can efficiently reduce interference
to the other cell despite the pilot contamination.

For the dense network in Fig.~\ref{Fig2}, we assume $M=7$ cells, $K=20$ users, and a larger inter-cell interference factor of $\rho=0.3$. In this case, for the considered range of $N_T$, the collaborative precoder designs are not able
to suppress inter-cell interference and AN leakage to other cells sufficiently well to outperform the selfish precoder designs.  In fact, for $N_T=400$, we obtain from (\ref{betazf})  $K_{{\rm CZF}>{\rm SZF}} \approx 16$, i.e.,
our analytical results suggest that the SZF precoder outperforms the CZF precoder for $K=20$ which is confirmed by Fig.~\ref{Fig2}. Nevertheless, for $N_T>400$, the ergodic secrecy rate for the CZF data precoder
will eventually surpass that for the SZF data precoder.
\subsection{Optimal Power Allocation}\label{s6b}
In this subsection, we investigate the dependence of the ergodic secrecy rate on the power allocation factor $\phi$ and study the impact of system parameters such as $\beta$, $M$, and $\rho$ on the optimal
$\phi$ that maximizes the ergodic secrecy rate. The results in this subsection were generated based on the analytical expressions in (\ref{secnk}), (\ref{Ceve}), and (\ref{srci-opt}) for the SRCI data precoder,
(\ref{crci-opt}) for the CRCI data precoder, and (\ref{rmfzf}) for the MF, SZF, and CZF data precoders.
\begin{figure}[t]
\centering
 \begin{minipage}[b]{0.45\linewidth} \hspace*{-1cm}
  \centering
  \includegraphics[width=3.7in]{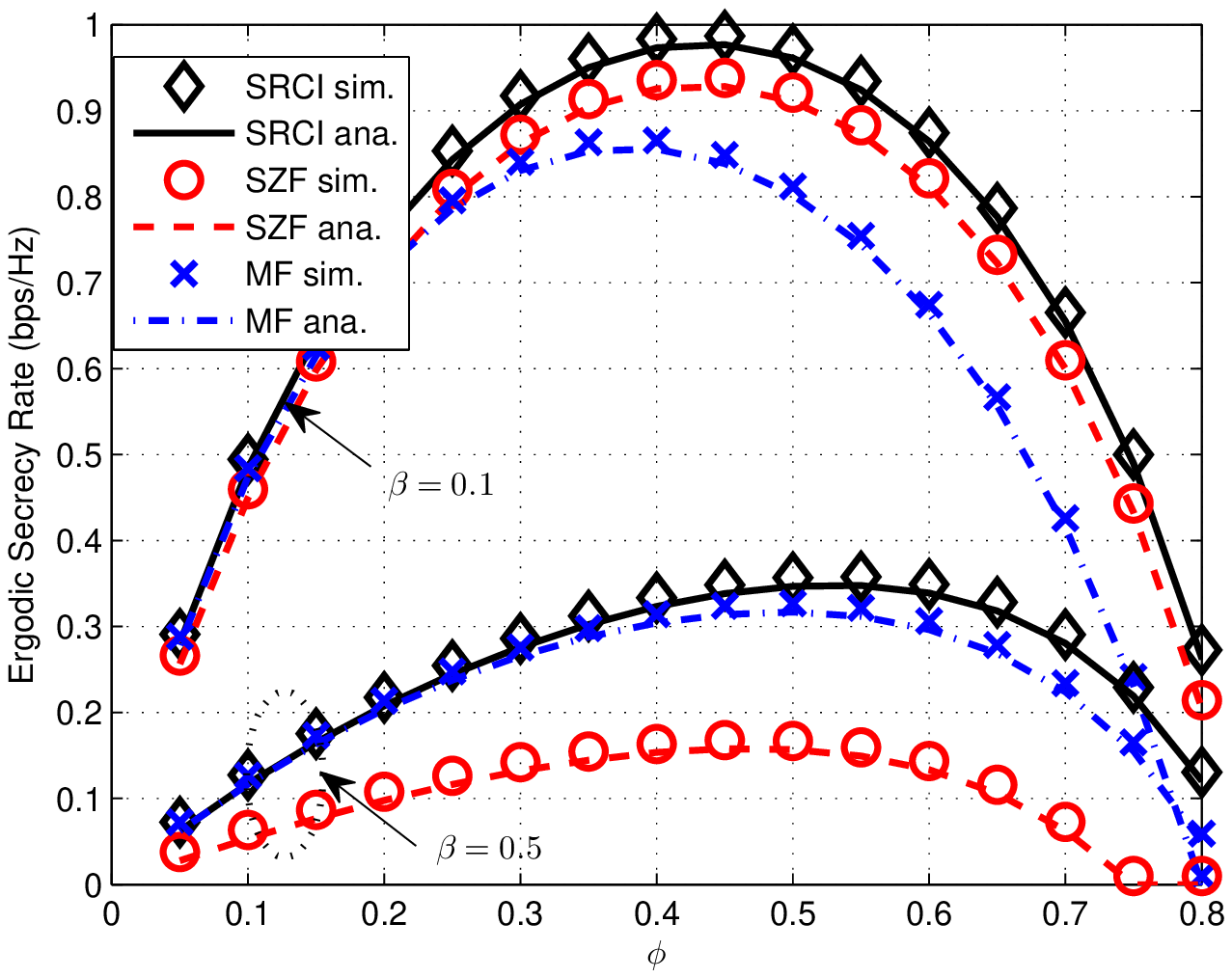}\vspace*{-8mm}\\
  \caption{Ergodic secrecy rate vs.~$\phi$ for different selfish data precoders for a network with $P_T=10$ dB, $N_T=100$, $p_\tau=P_T/K$, $\alpha=0.1$, $\rho=0.1$, and $M=7$. }\label{Fig3}
 \end{minipage}\hspace*{1.1cm}
 \begin{minipage}[b]{0.45\linewidth} \hspace*{-1cm}
  \centering
  \includegraphics[width=3.7in]{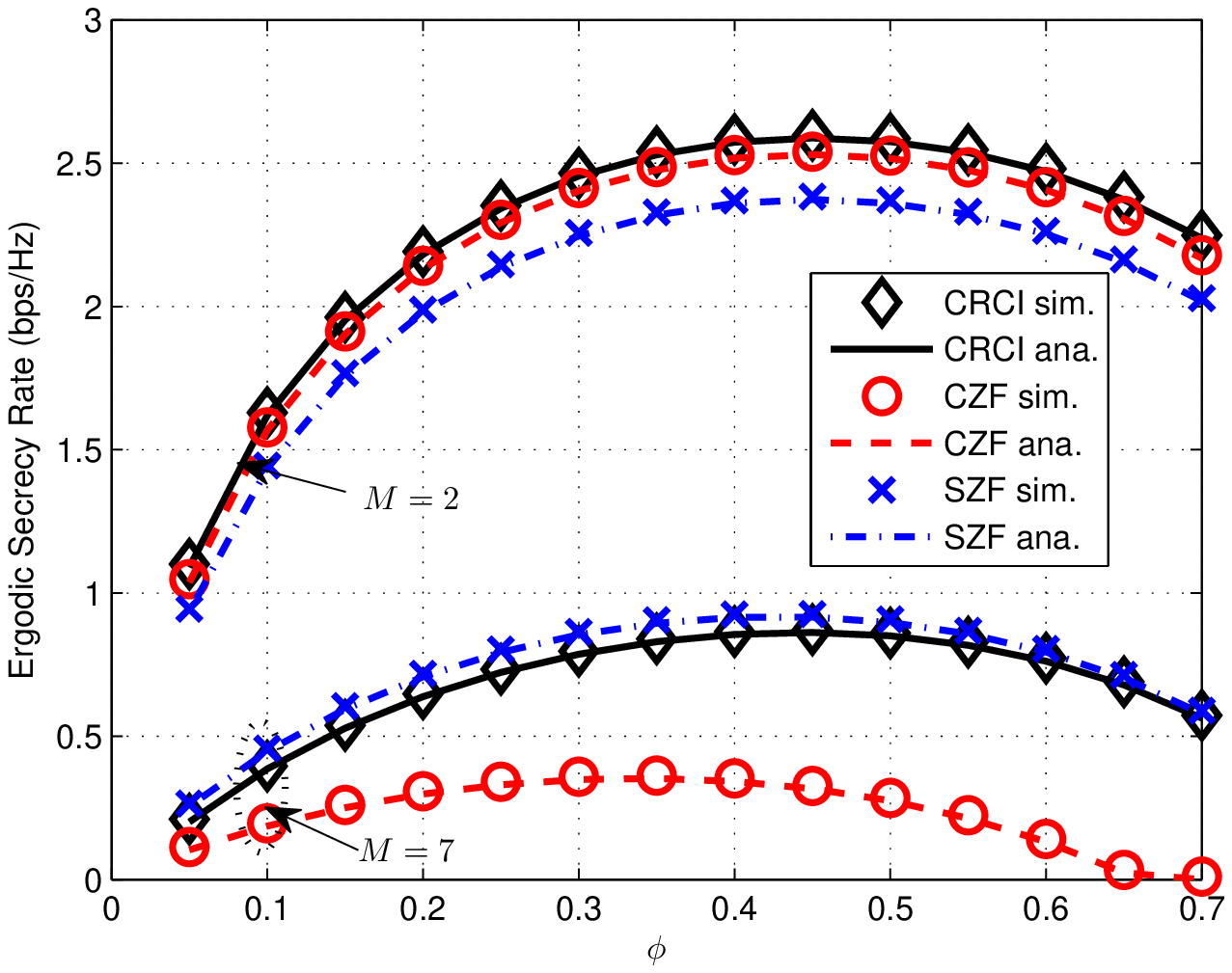}\vspace*{-8mm}\\
  \caption{Ergodic secrecy rate vs. $\phi$ for different data precoders for a network with $P_T=10$ dB, $N_T=100$, $p_\tau=P_T/K$, $\alpha=0.1$, $\beta=0.1$, and $\rho=0.1$. }\label{Fig4}
 \end{minipage}\hspace*{1.1cm}
  \end{figure}

Fig.~\ref{Fig3} depicts the ergodic secrecy rate of the $k^{\rm th}$ MT in the $n^{\rm th}$ cell for the selfish data precoders SRCI, SZF, and MF as a function of the power allocation factor $\phi$.
All curves are concave and have a single maximum. For $\phi=0$ only AN is transmitted, hence
$R_{nk}^{\rm sec}=0$ results since no data can be transmitted. For $\phi=1$, no AN is transmitted, hence $R_{nk}^{\rm sec}=0$ results since the capacity of the eavesdropper becomes unbounded
(recall that we make the worst-case assumption that the eavesdropper can receive noise-free). For $0<\phi<1$, a positive secrecy rate may result depending on the system parameters and the precoding
schemes. Since we keep the total transmit power fixed, the transmit power per MT decreases with increasing $\beta$. To compensate for this effect, the portion of the total transmit power allocated to data
transmission should increase. This is confirmed by Fig.~\ref{Fig3} where the optimal value of $\phi$ for $\beta = 0.5$ is larger than that for $\beta=0.1$. Furthermore, for a given $\beta$,
the optimal $\phi$ is the larger, the better the performance of the adopted data precoder is, i.e., for a more effective data precoder, transmitting the data signal with higher power is more beneficial, whereas
for a less effective data precoder impairing the eavesdropper with a higher AN power is more beneficial.
%

In Fig.~\ref{Fig4}, we show the ergodic secrecy rate vs.~$\phi$ for the CRCI, CZF, and SZF precoders. Similar to our observations in Fig.~\ref{Fig3}, for given system parameters, the optimal $\phi$ tends to be larger
for more effective precoders that achieve a better performance. For the system with $M=7$, this can be observed by comparing the optimal $\phi$ for the SZF and CZF precoders. Furthermore, while
for the smaller system with $M=2$ cells collaborative precoding is always preferable, for $M=7$, SZF precoding outperforms CZF and CRCI precoding for all considered values of $\phi$, as the collaborative designs
are not able to effectively suppress the interference and AN leakage to the $(M-1)K=60$ users in the other cells with the available $N_T=100$ antennas. In particular, from (\ref{betazf}), we obtain $K_{\rm CZF>SZF} \leq 18$
for $M=2$ and $K_{\rm CZF>SZF} \leq 5$ for $M=7$, which confirms the results shown in Fig.~\ref{Fig4}.
\begin{figure}[t]
\centering
 \begin{minipage}[b]{0.45\linewidth} \hspace*{-1cm}
  \centering
  \includegraphics[width=3.7in]{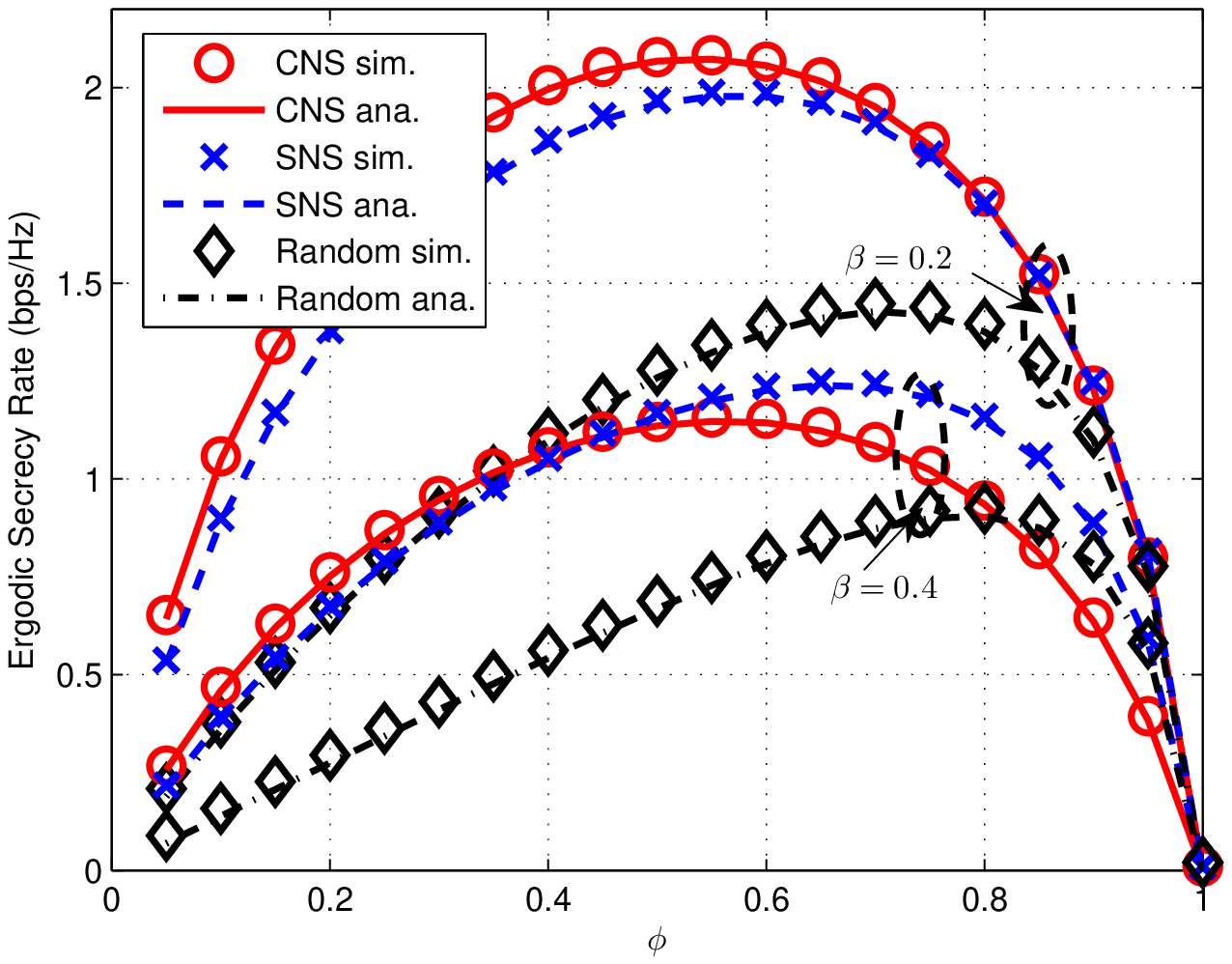}\vspace*{-8mm}\\
 \caption{Ergodic secrecy rate vs. $\phi$ for different AN precoders for a network with $P_T=10$ dB, $N_T=100$, $p_\tau=P_T/K$, $M=2$, $\rho=0.1$, and $\alpha=0.1$. }\label{Fig45}
 \end{minipage}\hspace*{1.1cm}
 \begin{minipage}[b]{0.45\linewidth} \hspace*{-1cm}
  \centering
  \includegraphics[width=3.7in]{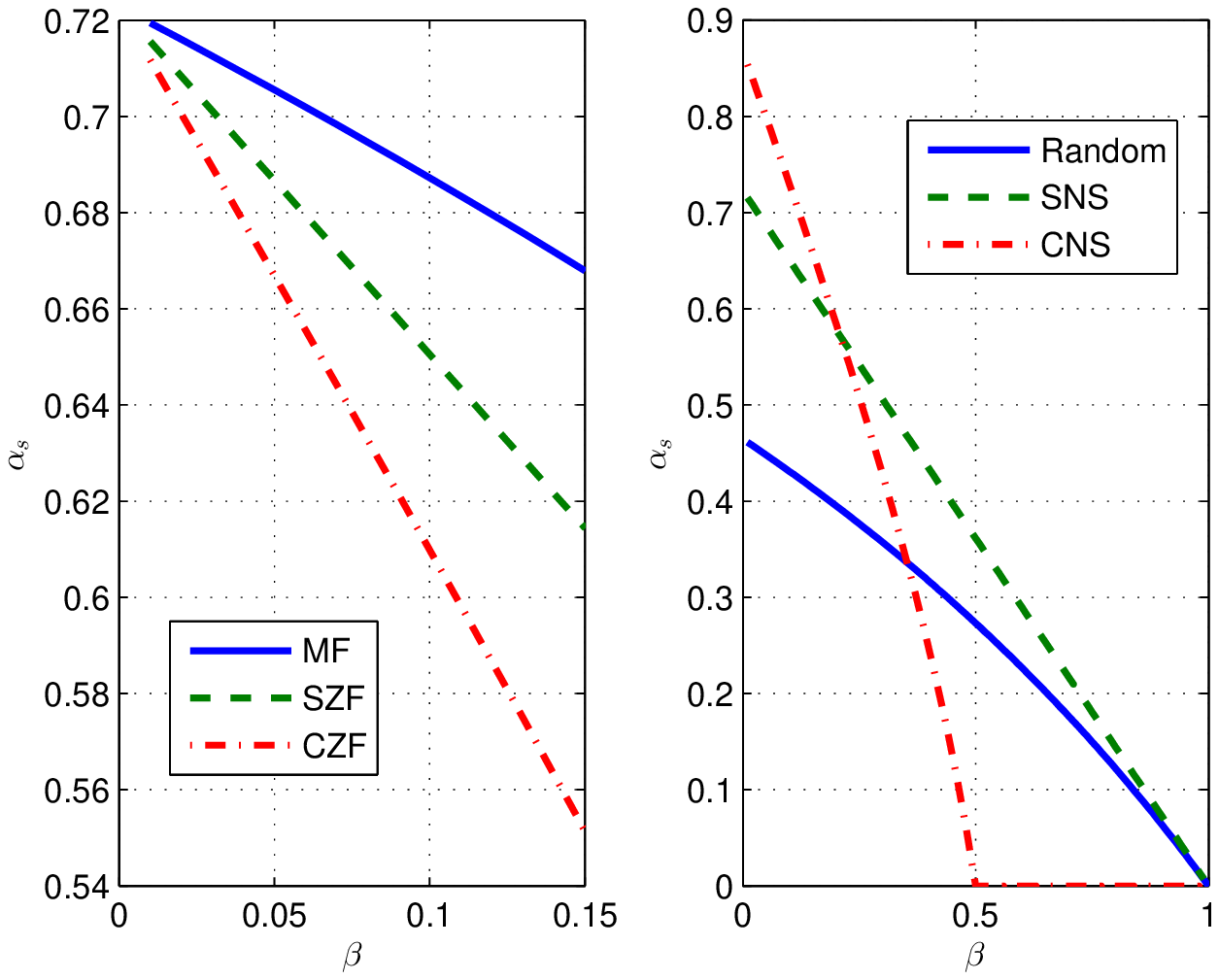}\vspace*{-8mm}\\
\caption{$\alpha_s$ vs. $\beta$ for different data and AN precoders for a network with $P_T=10$ dB, $N_T=100$, $p_\tau=P_T/K$, $\rho=0.3$, and $M=2$. }\label{Fig5}
 \end{minipage}\hspace*{1.1cm}
  \end{figure}
Fig.~\ref{Fig45} depicts the ergodic secrecy rate vs. $\phi$ for the considered conventional AN precoder structures. We consider a lightly loaded network with $\beta=0.2$ and a moderately loaded network with $\beta=0.4$. For $\beta=0.2$, the CNS
AN precoder outperforms the SNS AN precoder since, in this case, for the CNS AN precoder, the negative impact of having (slightly) fewer dimensions available for degrading the eavesdropper's channel (smaller value of $L$) is outweighed by the
positive impact of causing less AN leakage (smaller value of $\tilde Q$). On the other hand, for $\beta=0.4$, the CNS AN precoder has a substantially smaller $L$ than the SNS precoder which cannot be compensated by its larger $\tilde Q$. Despite
having the largest value of $L$, the random AN precoder has the worst performance for both considered cases because of its large AN leakage.

\subsection{Conditions for Non-Zero Secrecy Rate}\label{s6c}
In Section \ref{s5c}, we showed that a positive ergodic secrecy rate is possible only if $\alpha<\alpha_s$. In Fig.~\ref{Fig5}, using (\ref{alphas}), we plot $\alpha_s$ as a function of $\beta$. In the
left hand side subfigure, we compare MF, SZF, and CZF data precoding for SNS AN precoding, and in the right hand side subfigure, we compare random, SNS, and CNS AN precoding for SZF data precoding.
The comparison of the data precoders reveals that although SZF and CZF entail a much higher complexity, MF precoding achieves a larger $\alpha_s$. Therefore, if the eavesdropper has a large number
of antennas and small ergodic secrecy rates are targeted, simple MF precoding is always preferable. On the other hand, whether SNS or CNS AN precoder is preferable depends on the system load.
For small values of $\beta$, CNS AN precoding can tolerate more eavesdropper antennas, whereas for large values of $\beta$, SNS AN precoding is preferable. Random AN precoding is outperformed
by SNS AN precoding for any value of $\beta$. A closer examination of (\ref{alphas}) reveals that this is always true if S/CZF data precoders are employed. However, for the MF data precoder, there are parameter
combination for which random AN precoding outperforms SNS and CNS AN precoding.
\subsection{Low-Complexity POLY Data and AN Precoders}\label{s6d}
\begin{figure}[t]
\centering
 \begin{minipage}[b]{0.45\linewidth} \hspace*{-1cm}
  \centering
  \includegraphics[width=3.7in]{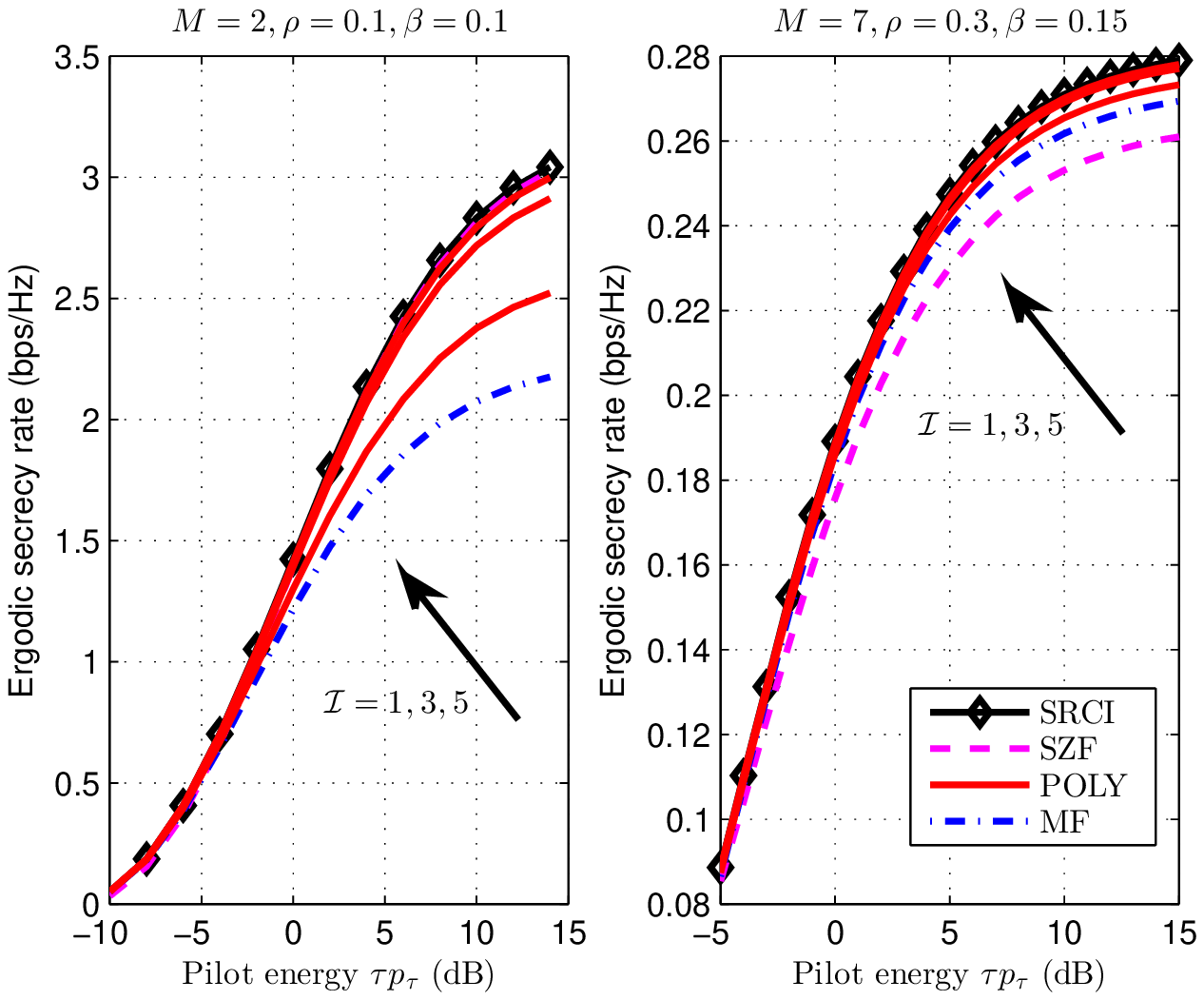}\vspace*{-8mm}\\
 \caption{Ergodic secrecy rate for POLY and conventional selfish data precoders for a network employing the optimal $\phi$, $P_T=10$ dB, $N_T=200$, and $\alpha=0.1$. }\label{Fig6}
 \end{minipage}\hspace*{1.1cm}
 \begin{minipage}[b]{0.45\linewidth} \hspace*{-1cm}
  \centering
  \includegraphics[width=3.7in]{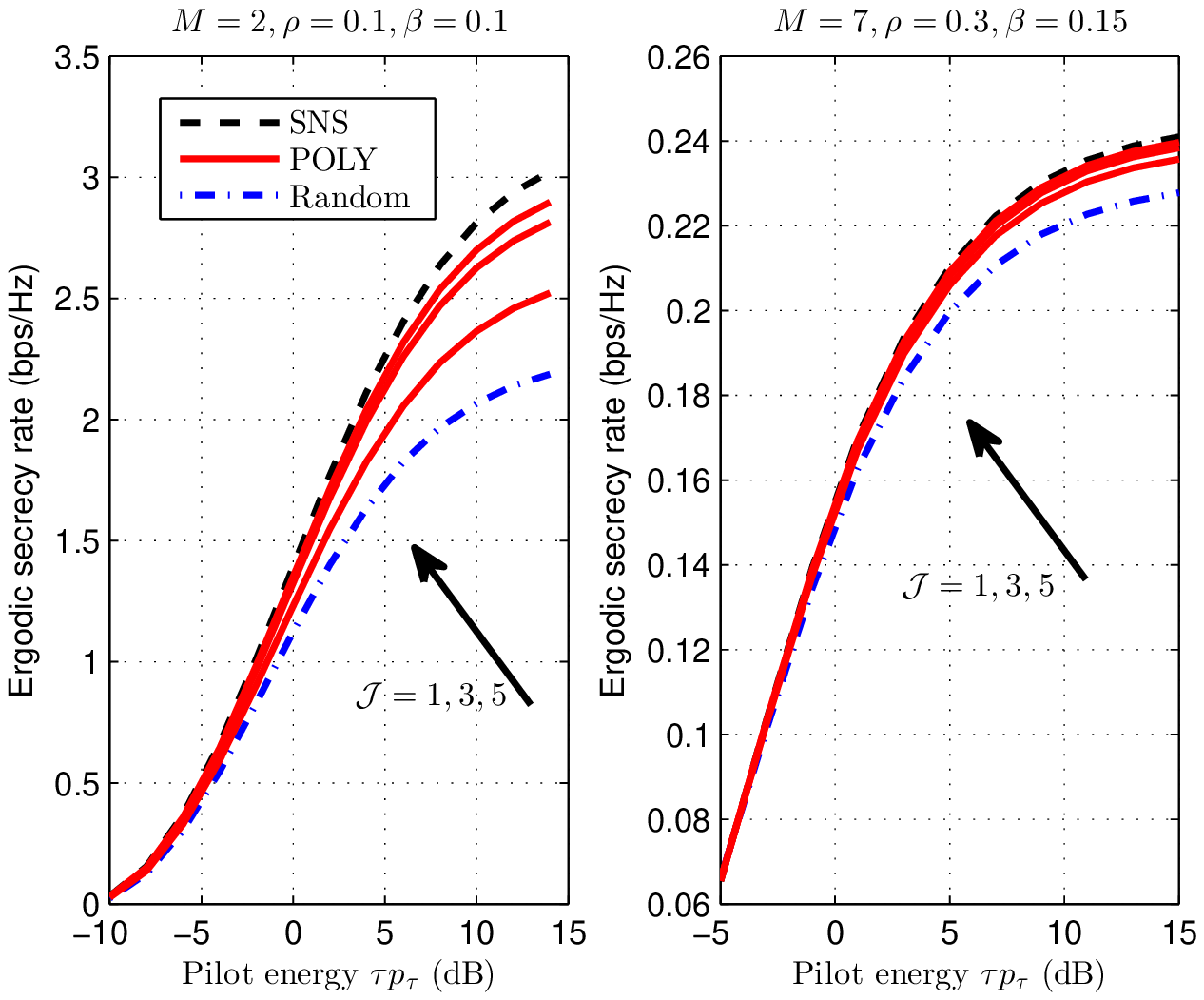}\vspace*{-8mm}\\
\caption{Ergodic secrecy rate for POLY, SNS, and random AN precoders for a network employing the optimal $\phi$, $P_T=10$ dB, $N_T=200$, and $\alpha=0.1$. }\label{Fig7}
 \end{minipage}\hspace*{1.1cm}
  \end{figure}

In this subsection, we evaluate the ergodic secrecy rates of the proposed low-complexity POLY data and AN precoders. To this end, we consider again a lightly loaded network with little inter-cell
interference ($M=2$, $\beta=0.1$, $\rho=0.1$) and a dense network with more inter-cell interference ($M=7$, $\beta=0.15$, $\rho=0.3$). All results shown in this section were obtained by simulation. For each simulation
point, the optimal value of $\phi$ was found numerically and applied. {In Figs.~\ref{Fig6} and \ref{Fig7}, we show the ergodic secrecy rate of the $k^{\rm th}$ MT in the $n^{\rm th}$ cell as a function of the
pilot energy, $\tau p_\tau$. As expected, for all considered schemes, the ergodic secrecy rate is monotonically increasing in the pilot energy since more accurate channel estimates improve the performance.}

In Fig.~\ref{Fig6}, we depict the ergodic secrecy rates for the proposed POLY data precoder for different values of ${\cal I}$ and compare them to those of conventional selfish
data precoders. For the sake of comparison, all data precoders are combined with the SNS AN precoder. As the number of terms of the polynomial ${\cal I}$ increase, the performance of the POLY data
precoder quickly improves and approaches that of the SRCI data precoder. The convergence is faster for the dense network considered in the right hand side subfigure, where the performance difference between
all precoders is smaller in general since interference cannot be as efficiently avoided as for the lightly loaded network.


In Fig.~\ref{Fig7}, we show the ergodic secrecy rates for the proposed POLY AN precoder for different values of ${\cal J}$ and compare them to those of the random and SNS
AN precoders. For the sake of comparison, all AN precoders are combined with SZF data precoding. The POLY AN precoder quickly approaches the performance of the SNS AN precoder as the polynomial order
${\cal J}$ increases. Similar to the POLY data precoders, the convergence is faster for the dense network where the performance differences between different AN precoders are also smaller. For
the denser network, even the random AN precoder is a viable option and suffers only from a small loss in performance compared to the SNS AN precoder.
\subsection{Complexity-Performance Tradeoff}
\begin{figure}[t]
\centering
 \begin{minipage}[b]{0.45\linewidth} \hspace*{-1cm}
  \centering
  \includegraphics[width=3.7in]{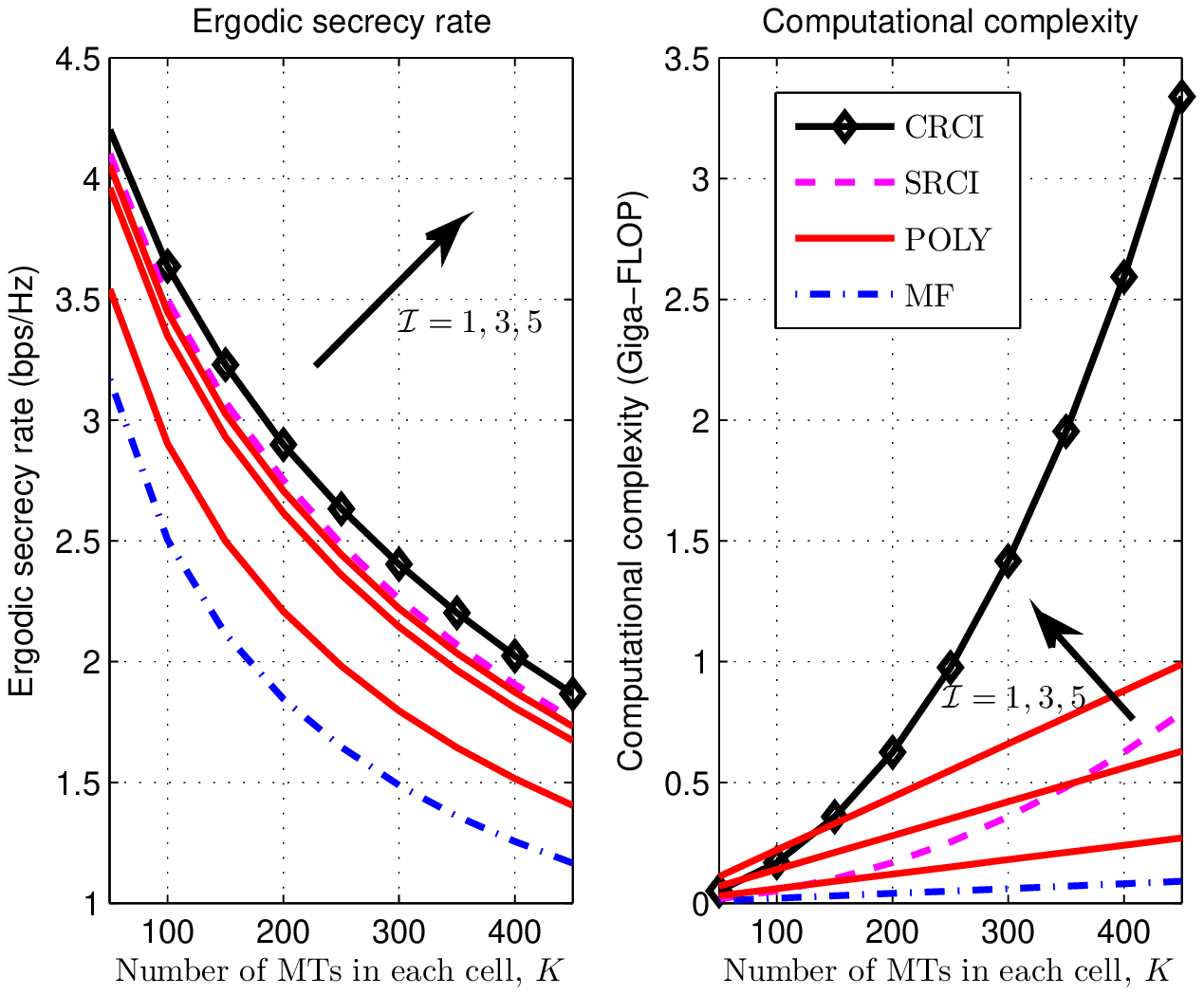}\vspace*{-8mm}\\
 \caption{Ergodic secrecy rate (left hand side) and computational complexity (right hand side) of various linear data precoders for a network employing $P_T=10$ dB, $N_T=1000$, $p_{\tau}=P_T/K$, $M=2$, $\rho=0.1$,
 $T-\tau=100$, and an SNS AN precoder. }\label{Fig8}
 \end{minipage}\hspace*{1.1cm}
 \begin{minipage}[b]{0.45\linewidth} \hspace*{-1cm}
  \centering
  \includegraphics[width=3.7in]{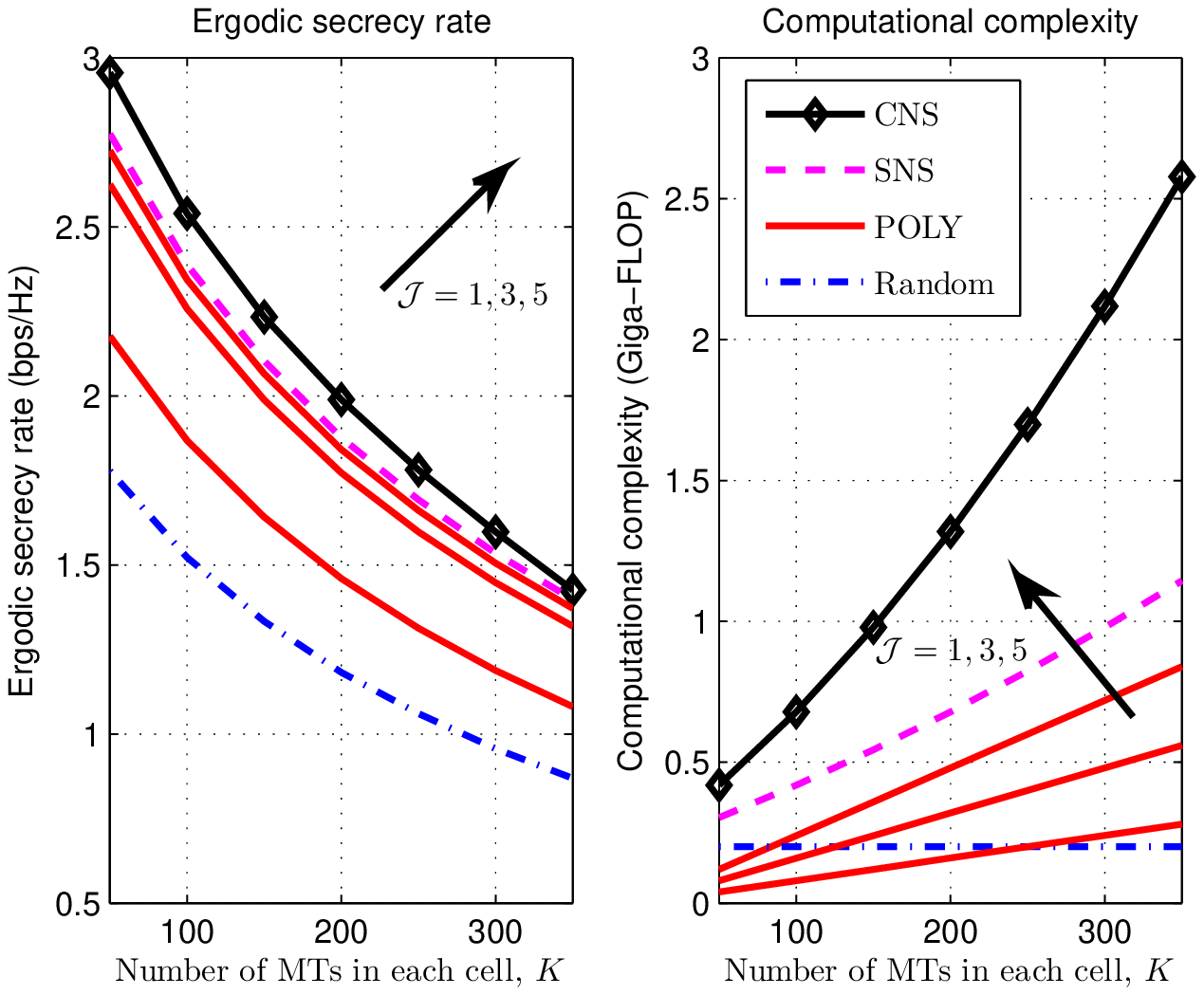}\vspace*{-8mm}\\
\caption{Ergodic secrecy rate (left hand side) and computational complexity (right hand side) of various linear AN precoders for a network employing $P_T=10$ dB, $N_T=1000$, $p_{\tau}=P_T/K$, $M=2$, $\rho=0.1$,
 $T-\tau=100$, and an SZF data precoder.}\label{Fig9}
 \end{minipage}\hspace*{1.1cm}
  \end{figure}
{In this subsection, we investigate the tradeoff between the ergodic secrecy rate performance and the computational complexity of the proposed data and AN precoders in Figs.~\ref{Fig8} and \ref{Fig9}, respectively.
In particular, Figs.~\ref{Fig8} and \ref{Fig9} depict the ergodic secrecy rate on the left hand side and the computational complexity (in Giga FLOP) on the right hand side, both as a function of the numbers of users in a cell. For the considered
setting, the performance gains of collaborative data and AN precoding compared to selfish strategies are moderate, but the associated increase in complexity is substantial, especially for large $K$. }

{Fig.~\ref{Fig8} illustrates that for the considered setting a POLY data precoder with ${\cal I}=1$ achieves a better performance than the MF precoder but has substantially lower complexity than the SRCI precoder. For large ${\cal I}$,
the POLY data precoder has a lower complexity than the SRCI precoder for large $K$. However, even for small $K$, the POLY precoder may be preferable as it does not incur the stability issues that may arise in the implementation of
the large-scale matrix inversions required for the SRCI precoder.}

{Fig.~\ref{Fig9} shows that for the considered setting the proposed POLY AN precoder with ${\cal J}=1$ outperforms the random AN precoder. The POLY AN precoder with ${\cal J}=5$ achieves almost the same performance as the SNS AN precoder but with a substantially lower complexity. We further observe that for small $K$, because of its efficient implementation via Horner's scheme, cf.~(\ref{Ahorner}), the proposed POLY AN precoder requires an even lower complexity than the random
AN precoder.}
\section{Conclusion \label{s7}}
In this paper, we considered downlink multi-cell massive MIMO systems employing linear data and AN precoding for physical layer security provisioning. We analyzed and compared the achievable ergodic secrecy rate of various
conventional data and AN precoders in the presence of pilot contamination. To this end, we also optimized the regularization constants of the selfish and collaborative RCI precoders in the presence of AN and multi-cell interference.
In addition, we derived linear POLY data and AN precoders which offer a good compromise between complexity and performance in massive MIMO systems. Interesting findings of this paper include: 1) Collaborative data precoders
outperform selfish designs only in lightly loaded systems where a sufficient number of degrees of freedom for suppressing inter-cell interference and sufficient resources for training are available. 2) Similarly, CNS AN precoding is preferable
over SNS AN precoding in lightly loaded systems as it causes less AN leakage to the information-carrying signal, whereas in more heavily loaded systems, CNS AN precoding does not have sufficient degrees of freedom for effectively
degrading the eavesdropper channel and SNS AN precoding is preferable. 3) For a large number of eavesdropper antennas, where only small positive secrecy rates are achievable, MF data precoding is always preferable compared
to SZF and CZF data precoding. 4) The proposed POLY data and AN precoders approach the performances of the SRCI data and SNS AN precoders with only a few terms in the respective matrix polynomials and are attractive options
for practical implementation.
\section*{Appendix}
\subsection{Proof of \textit{Proposition 1}}
Considering (\ref{eq3}) and (\ref{srci}), the effective signal power, i.e., the numerator in (\ref{Rnk}), can be expressed as \cite{mmse}
\begin{equation}
\label{sp}
\mathbb{E}^2[{\bf h}^k_{nn} {\bf f}_{nk}] = \gamma^2_1 \mathbb{E}^2[{\bf h}^k_{nn} {\bf L}_{nn} (\hat{\bf h}^k_{nn})^H]=\gamma^2_1 \mathbb{E}^2\bigg[\frac{{\bf h}^k_{nn} {\bf L}_{n,k} (\hat{\bf h}^k_{nn})^H}{1+\hat{\bf h}^k_{nn} {\bf L}_{n,k}
(\hat{\bf h}^k_{nn})^H}\bigg] = \frac{\gamma^2_1 (X_{nk}+A_{nk})^2}{(1+X_{nk})^2},
\end{equation}
where ${\bf L}_{n,k}=(\hat{\bf H}_{nn} \hat{\bf H}^H_{nn}-(\hat{\bf h}^k_{nn})^H\hat{\bf h}^k_{nn}+\kappa_1 {\bf I}_{N_T})^{-1}$, $X_{nk}= \mathbb{E}[\hat{\bf h}^k_{nn}{\bf L}_{n,k}(\hat{\bf h}^k_{nn})^H]$, and
$A_{nk} = \mathbb{E}[\tilde{\bf h}_{nn}^k {\bf L}_{n,k}(\hat{\bf h}^k_{nn})^H]$. On the other hand, the intra-cell interference term in the denominator of (\ref{Rnk}) can be expressed as
\begin{equation}
\mathbb{E}\bigg[\sum_{l \neq k}|{\bf h}^k_{nn} {\bf f}_{nl}|^2\bigg] = \gamma^2_1 \mathbb{E}\bigg[\frac{{\bf h}^k_{nn} {\bf L}_{n,k} \hat{\bf H}^H_{n,k} \hat{\bf H}_{n,k} {\bf L}_{n,k} ({\bf h}^k_{nn})^H}{\left(1+\hat{\bf h}^k_{nn} {\bf L}_{n,k}
(\hat{\bf h}^k_{nn})^H\right)^2}\bigg]=\frac{\gamma^2_1 (Y_{nk}+B_{nk})}{(1+X_{nk})^2},
\end{equation}
where ${\hat{\bf H}}_{n,k}$ is equal to $\hat{{\bf H}}_{nn}$ with the $k^{\rm th}$ row removed, and $Y_{nk}=\mathbb{E}[\hat{\bf h}^k_{nn} {\bf L}_{n,k} \hat{\bf H}^H_{n,k} \hat{\bf H}_{n,k} {\bf L}_{n,k} (\hat{\bf h}^k_{nn})^H]$ and $B_{nk}=\mathbb{E}[
\tilde{\bf h}_{nm}^k {\bf L}_{n,k} \hat{\bf H}^H_{n,k} \hat{\bf H}_{n,k} {\bf L}_{n,k} (\tilde{\bf h}^k_{nn})^H]$.

Due to pilot contamination, the data precoding matrix of the $m^{\rm th}$ BS is a function of the channel vectors between the $m^{\rm th}$
BS and the $k^{\rm th}$ MTs in all cells. Hence, the inter-cell interference from the BSs in adjacent cells is obtained as
\begin{equation}
\label{eq13a}
\mathbb{E}[|{\bf h}^k_{mn} {\bf f}_{mk}|^2]=\frac{\gamma^2_1 (X_{nk}+A_{nk})^2}{(1+X_{nk})^2}+\frac{1+p_\tau \tau \sum_{l \neq m}^M \beta^k_{ml}}{1+p_\tau \tau \sum_{l=1}^M \beta^k_{ml}}.
\end{equation}

Meanwhile, by exploiting (\ref{sp}), (\ref{eq13a}), and the definition of the variance, i.e., ${\rm var}[x]=\mathbb{E}[x^2]-\mathbb{E}^2[x]$, we obtain for the first term of the denominator of (\ref{Rnk}), ${\rm var}[{\bf h}^k_{nn} {\bf f}_{nk}]=\frac{1+p_\tau \tau \sum_{m \neq n}^M \beta^k_{nm}}{1+p_\tau \tau \sum_{m=1}^M \beta^k_{nm}}$. According to \cite[Eq. (16)]{mmse} and \cite[Theorem 7]{Evans}, for $N_T\to\infty$ and constant $\beta$, $X_{nk}$ converges to $\mathcal{G}(\beta,\kappa_1)$ defined in (\ref{g}) and $A_{nk} \to 0$. Similarly, $Y_{nk}$ and $B_{nk}$ approach
\begin{equation}
Y_{nk}\stackrel{N_T\to\infty}{=}\mathcal{G}(\beta,\kappa_1)+\kappa_1 \frac{\partial}{\partial \kappa_1} \mathcal{G}(\beta,\kappa_1)
\end{equation}
and
\begin{equation}
B_{nk}\stackrel{N_T\to\infty}{=}\frac{\vartheta_{nk}}{\theta_{nk}} (1+\mathcal{G}(\beta,\kappa_1))^2 \left(\mathcal{G}(\beta,\kappa_1)+\kappa_1 \frac{\partial}{\partial \kappa_1} \mathcal{G}(\beta,\kappa_1)\right),
\end{equation}
respectively, where $\frac{\partial}{\partial \kappa_1} \mathcal{G}(\beta,\kappa_1)=-\frac{\mathcal{G}(\beta,\kappa_1) (1+\mathcal{G}(\beta,\kappa_1))^2}{\beta+\kappa_1 (1+\mathcal{G}(\beta,\kappa_1))^2}$.

Moreover, the inter-cell interference from other MTs (i.e., not the $k^{\rm th}$ MTs) is calculated as
\begin{equation}
\label{io}
\mathbb{E}\bigg[{\bf h}^k_{mn} {\bf F}_{m,k}{\bf F}^H_{m,k}({\bf h}^k_{mn})^H\bigg]=\mathbb{E}\bigg[{\rm tr}\left\{{\bf F}_{m,k}{\bf F}^H_{m,k}\right\}\bigg]=K-1,
\end{equation}
where ${\bf F}_{m,k}$ is equal to ${\bf F}_{m}$ with the $k^{\rm th}$ column removed. The first equality in (\ref{io}) is due to the fact that the precoding matrix
for the other MTs (i.e., not the $k^{\rm th}$ MTs) in adjacent cells are independent of ${\bf h}^k_{mn}$ and \cite[Lemma 11]{muller}, while the second equality
holds for $N_T \to \infty$.

On the other hand, the constant scaling factor $\gamma_1$ for SRCI precoding is given by \cite[Eq. (22)]{mmse}
\begin{equation}\label{gamma12}
\gamma^2_1=\frac{1}{\mathcal{G}(\beta,\kappa_1)+\kappa_1 \frac{\partial}{\partial \kappa_1} \mathcal{G}(\beta,\kappa_1)}.
\end{equation}
Hence, employing (\ref{sp})-(\ref{gamma12}) in (\ref{Rnk}), the received SINR in (\ref{gammalmmseNpc}) is obtained, which completes the proof of \textit{Proposition 1}.
\subsection{Proof of \textit{Theorem 1}}
The objective function in  (\ref{opt1}) can be rewritten as
\begin{eqnarray}
\label{x1}
{\rm mse}_n &=& \varsigma^2 p \mathbb{E}\bigg[{\rm Tr}\bigg\{\sum_{i=0}^{\cal I} \mu_i \bigg(\hat{\overline{\bf H}}_{nn} \hat{\overline{\bf H}}_{nn}^H\bigg)^{i+1} {\bf D}_{nn} \sum_{i=0}^{\cal I} \mu_i \bigg(\hat{\overline{\bf H}}_{nn} \hat{\overline{\bf H}}_{nn}^H\bigg)^{i+1}\bigg\}\bigg]
\nonumber\\
\nonumber &&+ \varsigma^2 p \mathbb{E}\bigg[{\rm Tr}\bigg\{\sum_{i=0}^{\cal I} \mu_i \bigg(\hat{\overline{\bf H}}_{nn} \hat{\overline{\bf H}}_{nn}^H\bigg)^i \hat{\overline{\bf H}}_{nn}\tilde{\bf \overline{H}}_{nn}^H {\bf D}_{nn} \tilde{\bf \overline{H}}_{nn} \hat{\overline{\bf H}}_{nn}^H \sum_{i=0}^{\cal I} \mu_i \bigg(\hat{\overline{\bf H}}_{nn} \hat{\overline{\bf H}}_{nn}^H\bigg)^i\bigg\}\bigg]\nonumber\\
&&-2 \varsigma \sqrt{p}\mathbb{E}\bigg[{\rm Tr}\bigg\{{\bf D}^{1/2}_{nn}\sum_{i=0}^{\cal I} \mu_i \bigg(\hat{\overline{\bf H}}_{nn} \hat{\overline{\bf H}}_{nn}^H\bigg)^{i+1}\bigg\}\bigg]+1+ \varsigma^2 P_{\rm AN}+\varsigma^2 {\rm Tr}\left\{\boldsymbol{\Sigma}_n\right\},
\end{eqnarray}
where we exploited $\mathbb{E}[{\bf s}_n {\bf s}^H_n]={\bf I}_K$, the definition of $P_{\rm AN}$ given in Theorem 1, the definition of ${\bf F}_n$ in (\ref{W}), the definition $\frac{1}{\sqrt{N_T}}{\bf H}_{nn}=\hat{{\bf \overline{H}}}_{nn}+\tilde{\bf \overline{H}}_{nn}$,
and $\tilde{\bf \overline{H}}_{nn}=\frac{1}{\sqrt{N_T}}\tilde{\bf {H}}_{nn} $.

In the following, we simplify the right hand side (RHS) of (\ref{x1}) term by term. To this end, we denote the first three terms on the RHS of (\ref{x1}) by $t_1$, $t_2$, and $t_3$, respectively.
{Using a result from free probability theory \cite{free}}, the first term converges to \cite[Theorem 1]{poly}
\begin{equation}
t_1 =\varsigma^2 p {\rm Tr}\left\{{\bf D}_{nn}\right\}\mathbb{E}\bigg[{\rm Tr}\bigg\{\bigg(\sum_{i=0}^{\cal I} \mu_i \bigg(\hat{\overline{\bf H}}_{nn} \hat{\overline{\bf H}}_{nn}^H\bigg)^{i+1}\bigg)^2\bigg\}\bigg],
\label{eq49}
\end{equation}
as matrix ${\bf D}_{nn}$ is free from $\sum_{i=0}^{\cal I} \mu_i \bigg(\hat{\overline{\bf H}}_{nn} \hat{\overline{\bf H}}_{nn}^H\bigg)^{i+1}$. Similarly, the third term converges to
\begin{equation}
t_3 =-2 \varsigma \sqrt{p} {\rm Tr}\left\{{\bf D}^{1/2}_{nn}\right\}\mathbb{E}\bigg[{\rm Tr}\bigg\{\sum_{i=0}^{\cal I} \mu_i \bigg(\hat{\overline{\bf H}}_{nn} \hat{\overline{\bf H}}_{nn}^H\bigg)^{i+1}\bigg\}\bigg].
\end{equation}
Furthermore, the second term can be rewritten as
\begin{eqnarray}
t_2 & \stackrel{{\rm (a)}}{=}& \varsigma^2 p \mathbb{E}\bigg[ {\rm Tr}\bigg\{\tilde{\bf \overline{H}}^H_{nn} {\bf D}_{nn} \tilde{\bf \overline{H}}_{nn}\bigg\} {\rm Tr}\bigg\{
\sum_{i=0}^{\cal I} \mu_i \bigg(\hat{\overline{\bf H}}_{nn} \hat{\overline{\bf H}}_{nn}^H\bigg)^{i} \hat{\overline{\bf H}}_{nn} \hat{\overline{\bf H}}_{nn}^H \sum_{i=0}^{\cal I} \mu_i \bigg(\hat{\overline{\bf H}}_{nn}
\hat{\overline{\bf H}}_{nn}^H\bigg)^{i}\bigg\}\bigg]\nonumber\\
& \stackrel{{\rm ({b})}}{=}&\varsigma^2 p N_T {\rm Tr}\left\{{\bf D}_{nn} \boldsymbol{\Delta}_n\right\},
\label{eq51}
\end{eqnarray}
where (a) follows again from \cite[Theorem 1]{poly} and ({b}) results from $\mathbb{E}[{\rm Tr}\{\tilde{\bf \overline{H}}^H_{nn} {\bf D}_{nn} \tilde{\bf \overline{H}}_{nn}\}]={\rm Tr}\left\{{\bf D}_{nn} \boldsymbol{\Delta}_n\right\}$, where
$\boldsymbol{\Delta}_n$ is defined in Theorem 1, (\ref{W}), and the constraint in (\ref{opt1}).

Exploiting (\ref{eq49})-(\ref{eq51}) and the eigen-decomposition of matrix $\hat{\overline{\bf H}}_{nn} \hat{\bf \overline{H}}^H_{nn}={\bf T} \boldsymbol{\Lambda} {\bf T}^H$, where diagonal matrix $\boldsymbol{\Lambda}=
{\rm diag}\left(\lambda_1,\ldots,\lambda_K\right)$ contains all eigenvalues and unitary matrix ${\bf T}$ contains the corresponding eigenvectors, the asymptotic average MSE becomes
\begin{eqnarray}
\label{eq53}
{\rm mse}_n &=&\mathbb{E}\bigg[\varsigma^2 p {\rm Tr}\left\{{\bf D}_{nn}\right\}    {\rm Tr}\bigg\{\boldsymbol{\Lambda}^2 \bigg(\sum_{i=0}^{\cal I} \mu_i \boldsymbol{\Lambda}^i\bigg)^2\bigg\}
- 2 \varsigma \sqrt{p}{\rm Tr}\left\{{\bf D}^{1/2}_{nn}\right\} {\rm Tr}\bigg\{\sum_{i=0}^{\cal I} \mu_i \boldsymbol{\Lambda}^{i+1}\bigg\}\bigg] \nonumber\\
&&+1+ \varsigma^2 P_{\rm AN}+\varsigma^2 {\rm Tr}\left\{\boldsymbol{\Sigma}_n\right\}+\varsigma^2pN_T {\rm Tr}\left\{{\bf D}_{nn} \boldsymbol{\Delta}_n\right\}.
\end{eqnarray}
Next, we introduce the Vandermonde matrix ${\bf C}_1 \in \mathbb{R}^{K \times ({\cal I}+1)}$, where $[{\bf C}_1]_{i,j}=\lambda^{j-1}_i$, and $\boldsymbol{\lambda}=[\lambda_1,\ldots,\lambda_K]^T$, which
allows us to rewrite (\ref{eq53}) in compact form as
\begin{eqnarray}
\label{eq54}
{\rm mse}_n &=&\lim_{K \to \infty} \frac{1}{K} \mathbb{E}\bigg[\varsigma^2 p  {\rm Tr}\left\{{\bf D}_{nn}\right\}  \boldsymbol{\mu}^T {\bf C}^T_1 \boldsymbol{\Lambda}^2  {\bf C}_1 \boldsymbol{\mu}
 - 2 \varsigma \sqrt{p} {\rm Tr}\left\{{\bf D}^{1/2}_{nn}\right\} \boldsymbol{\mu}^T {\bf C}^T_1 \boldsymbol{\lambda}\bigg]\nonumber\\
&& \qquad+1+ \varsigma^2 P_{\rm AN}+\varsigma^2 {\rm Tr}\left\{\boldsymbol{\Sigma}_n\right\} +\varsigma^2p N_T {\rm Tr}\left\{{\bf D}_{nn} \boldsymbol{\Delta}_n\right\}.
\end{eqnarray}
Similarly, the  constraint in (\ref{opt1}) can be expressed as
\begin{equation}
\lim_{K\to\infty}\frac{1}{K} \mathbb{E}\bigg[\boldsymbol{\mu}^T {\bf C}^T_1 \boldsymbol{\Lambda} {\bf C}_1 \boldsymbol{\mu}\bigg]=N_T.
\label{eq55}
\end{equation}
Thus, the Lagrangian function of primal problem (\ref{opt1}) can be expressed as $ {\cal L}_1(\boldsymbol{\mu},\varsigma) =  {\rm mse}_n +\epsilon_1 ( \lim_{K\to\infty}
\frac{1}{K} \mathbb{E} [\boldsymbol{\mu}^T {\bf C}^T_1 \boldsymbol{\Lambda} {\bf C}_1 \boldsymbol{\mu}]-N_T)$, where $\epsilon_1$ is the Lagrangian multiplier.
Taking the gradient of the Lagrangian function with respect to $\boldsymbol{\mu}$, and setting the result to zero, we obtain for the optimal coefficient vector $\boldsymbol{\mu}_{\rm opt}$:
\begin{equation}
\lim_{K \to \infty} \frac{1}{K} \mathbb{E}\left[{\bf C}^T_1 \boldsymbol{\Lambda}\left(\boldsymbol{\Lambda}+\frac{\epsilon_1}{{\rm Tr}\left\{{\bf D}_{nn}\right\} \varsigma^2  p} {\bf I}_K\right) {\bf C}_1\right] \boldsymbol{\mu} =
 \frac{{\rm Tr}\left\{{\bf D}^{1/2}_{nn}\right\}}{ \varsigma \sqrt{p}  {\rm Tr}\left\{{\bf D}_{nn}\right\}     }  \lim_{K \to \infty} \frac{1}{K} \mathbb{E}\left[{\bf C}^T_1 \boldsymbol{\lambda}\right].
\label{eq56}
\end{equation}
Furthermore, taking the derivative of $ {\cal L}_1(\boldsymbol{\mu},\varsigma)$ with respect to $\varsigma$ and equating it to zero, and multiplying both sides of (\ref{eq56}) by $\boldsymbol{\mu}^T$ and applying
(\ref{eq55}), we obtain
\begin{equation}
\label{eq57}
\frac{\epsilon_1}{\varsigma^2  p} =     {\rm Tr}\left\{{\bf D}_{nn} \boldsymbol{\Delta}_n\right\}  + \frac{P_{\rm AN}+{\rm Tr}\left\{\boldsymbol{\Sigma}_n\right\} }{  N_T p} .
\end{equation}
The expressions involving ${\bf C}_1$, $\boldsymbol{\Lambda}$, and $\boldsymbol{\lambda}$ in (\ref{eq56}) can be further simplified. For example, we obtain $\lim_{K \to \infty} \mathbb{E}\bigg[\frac{1}{K} \bigg[{\bf C}^T_1 \boldsymbol{\Lambda} {\bf C}_1\bigg]_{m,n}\bigg]=
\lim_{K \to \infty} \mathbb{E}\bigg[\frac{1}{K} \sum_{k=1}^K \lambda^{m+n-1}_k\bigg]$. Simplifying the other terms in (\ref{eq56}) in a similar manner and inserting (\ref{eq57}) into  (\ref{eq56}) we obtain the result in Theorem 1.
\subsection{Proof of \textit{Theorem 2}}
Exploiting $\mathbb{E}[{\bf z}_n {\bf z}^H_n]={\bf I}_{N_T}$, the constraint in (\ref{opt2}), and a similar approach as was used to arrive at (\ref{aaa}),
the objective function in (\ref{opt2}) can be simplified as $P_{\rm AN}=$
\begin{equation}
\label{eq73}
q\mathbb{E}\bigg[{\rm Tr}\left\{  {\bf G}_{nn} {\bf A}_n {\bf A}^H_n {\bf G}^H_{nn}\right\}\bigg]=q\mathbb{E}\bigg[{\rm Tr}\left\{ {\bf D}_{nn} \hat{{\bf H}}_{nn} {\bf A}_n {\bf A}^H_n \hat{\bf H}^H_{nn}\right\}\bigg]
+(1-\phi)P_T{\rm Tr}\{{\bf D}_{nn} \boldsymbol{\Delta}_n\}.
\end{equation}

Using (\ref{V}) and a similar approach as in Appendix B, (\ref{eq73}) can be rewritten as
\begin{eqnarray}
\label{y22}
P_{\rm AN} &=& (1-\phi)P_T{\rm Tr}\{{\bf D}_{nn} \boldsymbol{\Delta}_n\}\\
&&+q N_T{\rm Tr}\left\{{\bf D}_{nn}\right\}  \mathbb{E}\bigg[ -2 {\rm Tr}\bigg\{\sum_{j=0}^{\cal J} \nu_j \boldsymbol{\Lambda}^{j+2}\bigg\}+ {\rm Tr}\left\{\boldsymbol{\Lambda}\right\}+
{\rm Tr}\bigg\{\boldsymbol{\Lambda}\bigg(\sum_{i=0}^{\cal J} \nu_j \boldsymbol{\Lambda}^{j+1}\bigg)^2\bigg\}\bigg]\nonumber
\end{eqnarray}

Defining Vandermode matrix ${\bf C}_2 \in \mathbb{R}^{K \times ({\cal J}+1)}$, where $[{\bf C}_2]_{i,j}=\lambda_i^{j-1}$, we can rewrite (\ref{y22}) in compact form as $P_{\rm AN}=$
\begin{equation}
\label{y44}
 q N_T{\rm Tr}\left\{{\bf D}_{nn}\right\} \lim_{K \to \infty} \frac{1}{K} \mathbb{E}\bigg[-2 \boldsymbol{\nu}^T {\bf C}^T_2 \boldsymbol{\Lambda} \boldsymbol{\lambda}+ {\bf 1}^T
\boldsymbol{\lambda}+\boldsymbol{\nu}^T {\bf C}^T_2 \boldsymbol{\Lambda}^3 {\bf C}_2 \boldsymbol{\nu}\bigg]+(1-\phi)P_T{\rm Tr}\{{\bf D}_{nn} \boldsymbol{\Delta}_n\},
\end{equation}
where $ {\bf 1} $ denotes the all-ones column vector.
Taking into account the constraint in (\ref{opt2}), we can formulate the Lagrangian as  ${\cal L}_2(\boldsymbol{\nu}) = P_{\rm AN} +\epsilon_2(\lim_{K \to \infty}
\frac{1}{K}\mathbb{E}[\boldsymbol{\nu}^T {\bf C}^T_2 \boldsymbol{\Lambda}^2 {\bf C}_2 \boldsymbol{\nu}-2\boldsymbol{\nu}^T {\bf C}^T_2 \boldsymbol{\lambda}] +1)$ with Lagrangian multiplier $\epsilon_2$.
The optimal coefficient vector $\boldsymbol{\nu}_{\rm opt}$ is then obtained by taking the gradient of the Lagrangian function with respect to $\boldsymbol{\nu}$ and setting it to zero:
\begin{equation}\label{nuopt}
\lim_{K \to \infty} \mathbb{E}\bigg[{\bf C}^T_2 \boldsymbol{\Lambda}^2 \left(\boldsymbol{\Lambda}+\epsilon {\bf I}_K\right){\bf C}_2\bigg]\boldsymbol{\nu}=\lim_{K \to \infty} \mathbb{E}\bigg[{\bf C}^T_2 \left(\boldsymbol{\Lambda}+\epsilon {\bf I}_K\right) \boldsymbol{\lambda}\bigg],
\end{equation}
where we used $\epsilon=\frac{\epsilon_2}{q N_T{\rm Tr}\left\{{\bf D}_{nn}\right\}}$. Simplifying the terms in (\ref{nuopt}) by exploiting a similar approach as in Appendix B, we obtain the result in Theorem 2.

\end{document}